\pdfoutput=1   
\documentclass[usenatbib,usegraphicx,fl,n]{ctextemp_mn2e}
\usepackage{amsmath,amssymb,bm,color,multirow,multicol,array}
\usepackage{charter}   

\newcommand{\e}[1]{\times 10^{#1}}

\usepackage{ulem}

\def\bl{\mbox{\boldmath $l$}}
\def\bomega{{\mbox{\boldmath $\omega$}}}
\def\bmu{{\mbox{\boldmath $\mu$}}}

\def\bN{{\bf N}}

\def\be{\begin{equation}}
\def\ee{\end{equation}}
\def\ba{\begin{eqnarray}}
\def\ea{\end{eqnarray}}
\def\go{\mathrel{\raise.3ex\hbox{$>$}\mkern-14mu
             \lower0.6ex\hbox{$\sim$}}}
\def\lo{\mathrel{\raise.3ex\hbox{$<$}\mkern-14mu
             \lower0.6ex\hbox{$\sim$}}}

\def\bN{{\bf N}}

\title[3D MHD Simulations of Accretion onto Stars with Tilted Magnetic and Rotational Axes]{3D MHD Simulations of Accretion onto Stars with Tilted Magnetic and Rotational Axes}

\author[Romanova et al.]{\parbox{\textwidth}{M. M.~Romanova$^{1,2}$\thanks{E-mail of
corresponding author: \texttt{romanova@astro.cornell.edu}},   A. V.~Koldoba$^{3}$, G. V.~Ustyugova$^{4}$,
A. A. Blinova$^{1,2}$,  D. Lai$^{1,2}$, R. V. E.~Lovelace$^{1,2,5}$
}\vspace{0.4cm}\\
\parbox{\textwidth}{ 
$^{1}$Department of Astronomy, Cornell University, Ithaca, NY 14853-6801\\
$^{2}$Carl Sagan Institute, Cornell University, Ithaca, NY 14853-6801\\
$^{3}$Moscow Institute of Physics and Technology, Dolgoprudny, Moscow Region, 141700, Russia \\
$^{4}$Keldysh Institute for Applied Mathematics, Moscow, 125047,
Russia \\
$^{5}$Department of Applied and Engineering Physics, Cornell
University, Ithaca, NY 14853-6801}}

\begin{document}

\maketitle

\begin{abstract}

\noindent  We present results of global three-dimensional
(3D) magnetohydrodynamic (MHD) simulations of accretion onto 
magnetized stars where both the magnetic and rotational axes of the star are
tilted about the rotational axis of the disc. We observed that 
  initially the inner parts of the disc are warped, tilted, and precess
 due to the magnetic interaction between the 
magnetosphere and the disc. Later, larger tilted discs form with the size increasing
 with the magnetic moment of the star.
The normal vector to the discs are tilted at different angles, from 
 $\sim 5^\circ-10^\circ$ up to  $\sim 30^\circ-40^\circ$. 
Small tilts may result from the winding of the magnetic field lines about the rotational axis of the star and the action of the magnetic force
which tends to align the disc. 
Another possible explanation is the magnetic Bardeen-Petterson effect in which
the disc settles in the equatorial plane of the star due to 
 precessional and viscous torques in the disc. 
Tilted discs slowly precess with the time scale of the order of $\sim 50$ Keplerian 
periods at the reference radius ($\sim 3$ stellar radii).  
Our results can be applied to different types of stars where evidence of tilted discs and/or slow precession has been observed.

\end{abstract}


\section{Introduction}

Different types of disc-accreting stars have strong magnetic fields,
such as young T Tauri stars (e.g., \citealt{BouvierEtAl2007}),
accreting X-ray pulsars (e.g., \citealt{Vanderklis2006}), and  white dwarfs
(intermediate polars, e.g., \citealt{Warner1995,Warner2004,Hellier2001}).
 The magnetospheres of these stars
open magnetospheric gaps in the  surrounding accretion discs, giving rise to complex paths of accretion onto the star.
Many observational properties of these stars are determined by
the disc-magnetosphere interactions.

The magnetic field of stars
may be complex (e.g.,  \citealt{Johns-Krull2007}). However, at large distances, the dipole
component often dominates and is responsible for the
disc-magnetosphere interaction (e.g., \citealt{LongEtAl2007,LongEtAl2008,Gregory2011}). 
Spectropolarimetric observations show that in many young stars, the dipole component
of the field is tilted about the rotational axis of the star by an angle, $\theta\sim 10^\circ-20^\circ$ (e.g., \citealt{DonatiEtAl2007,DonatiEtAl2010,DonatiEtAl2011}). In general, the rotational axis of the star can also be tilted 
with respect to the rotational axis of the disc. 
Such misalignments may result from the varying 
angular momentum directions of the gas that falls onto the disc, as expected in the assembly of protoplanetary discs.

\begin{figure}
\centering
\includegraphics[width=8cm]{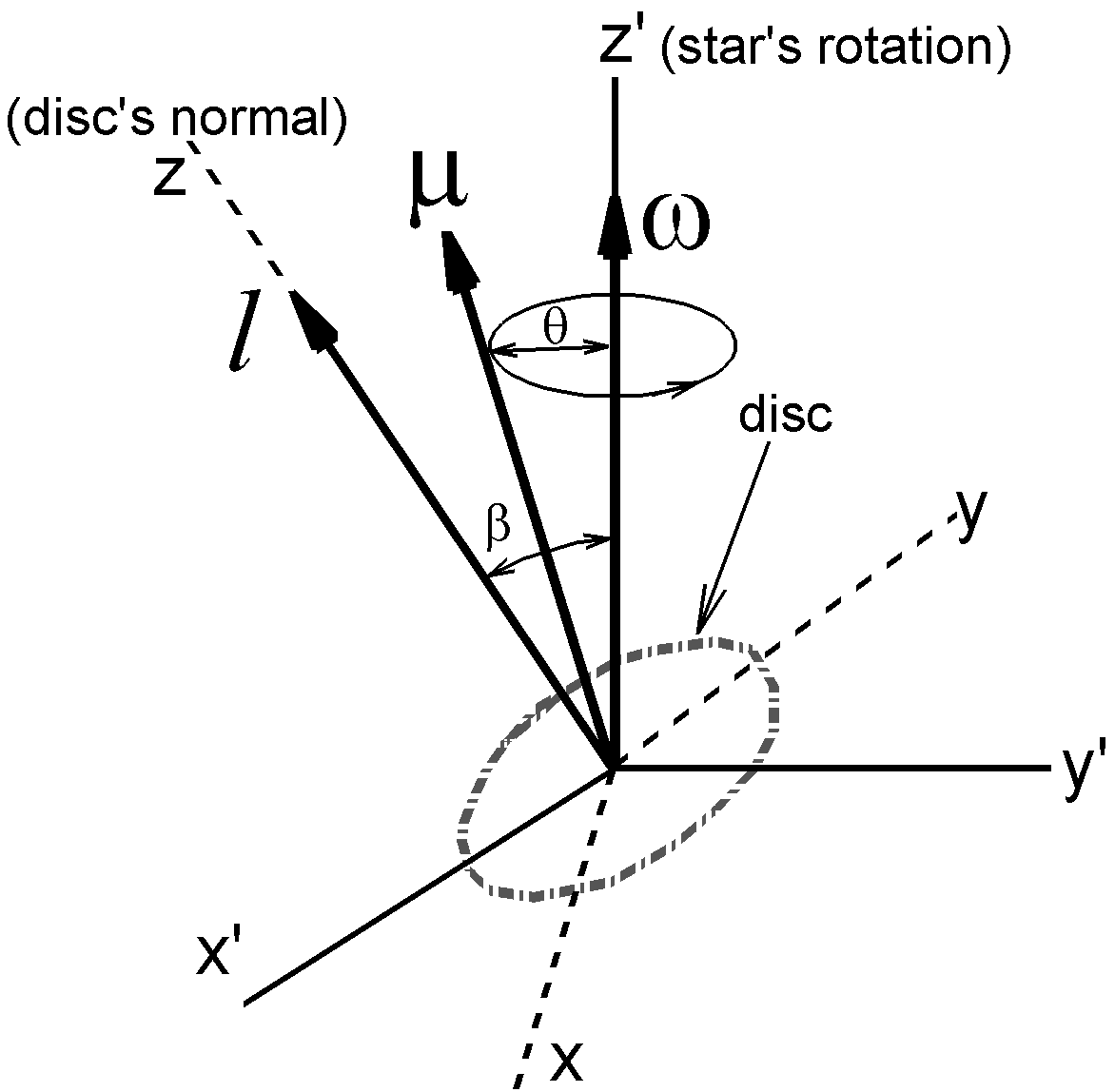}
\caption{Sketch shows coordinate systems used in the paper.
The coordinate system $x'y'z'$ is used in numerical simulations.
In this system, 
the $z'-$axis
is along the direction of the star's angular velocity vector,
${\bomega}$.  
The stellar dipole moment, $\bmu$ 
is tilted about  $\bomega$ by angle $\theta$. 
The local disc's angular momentum vector, $\bl$
is directed along the $z-$axis and is tilted by an angle $\beta$ with respect of ${\bomega}$. 
Initially, at $t=0$, the disc is located at the $xy-$plane. The coordinate system $xyz$ is used 
in our 3D plots.} \label{sketch}
\end{figure}

\begin{figure*}
\centering
\includegraphics[width=16cm]{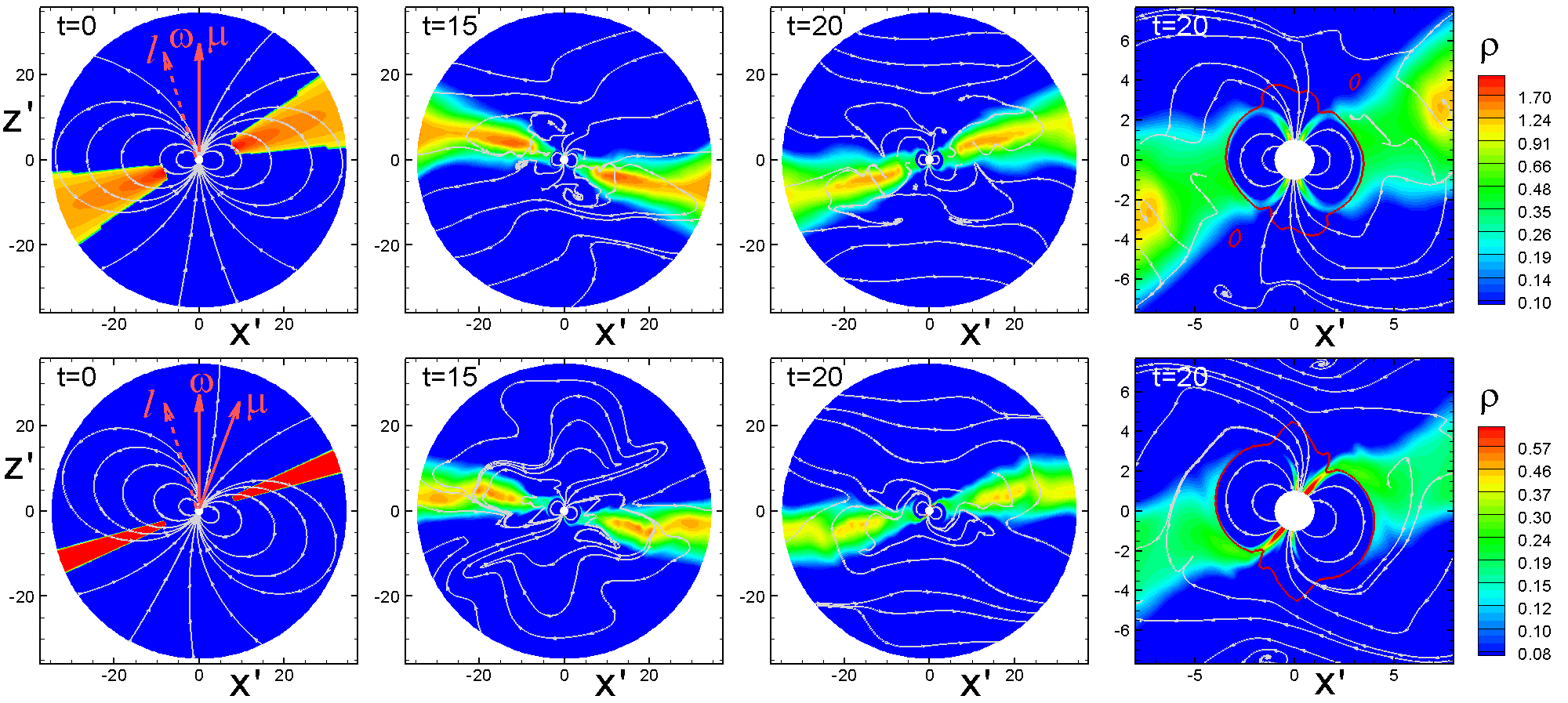}
\caption{$x'z'-$slices of density distribution in models A and B (top and bottom panels, respectively) at $t=0$,  $t=15$, and $t=20$. Top and bottom right panels show close view of matter flow near the magnetosphere at $t=20$. 
White lines show sample poloidal magnetic field lines. Red lines show the $\beta_1=1$ line. In this and other 2D plots we use the coordinate system $x'y'z'$ (see Fig. \ref{sketch}).}
\label{2d-initial-8}
\end{figure*}

\begin{figure}
\centering
\includegraphics[width=8cm]{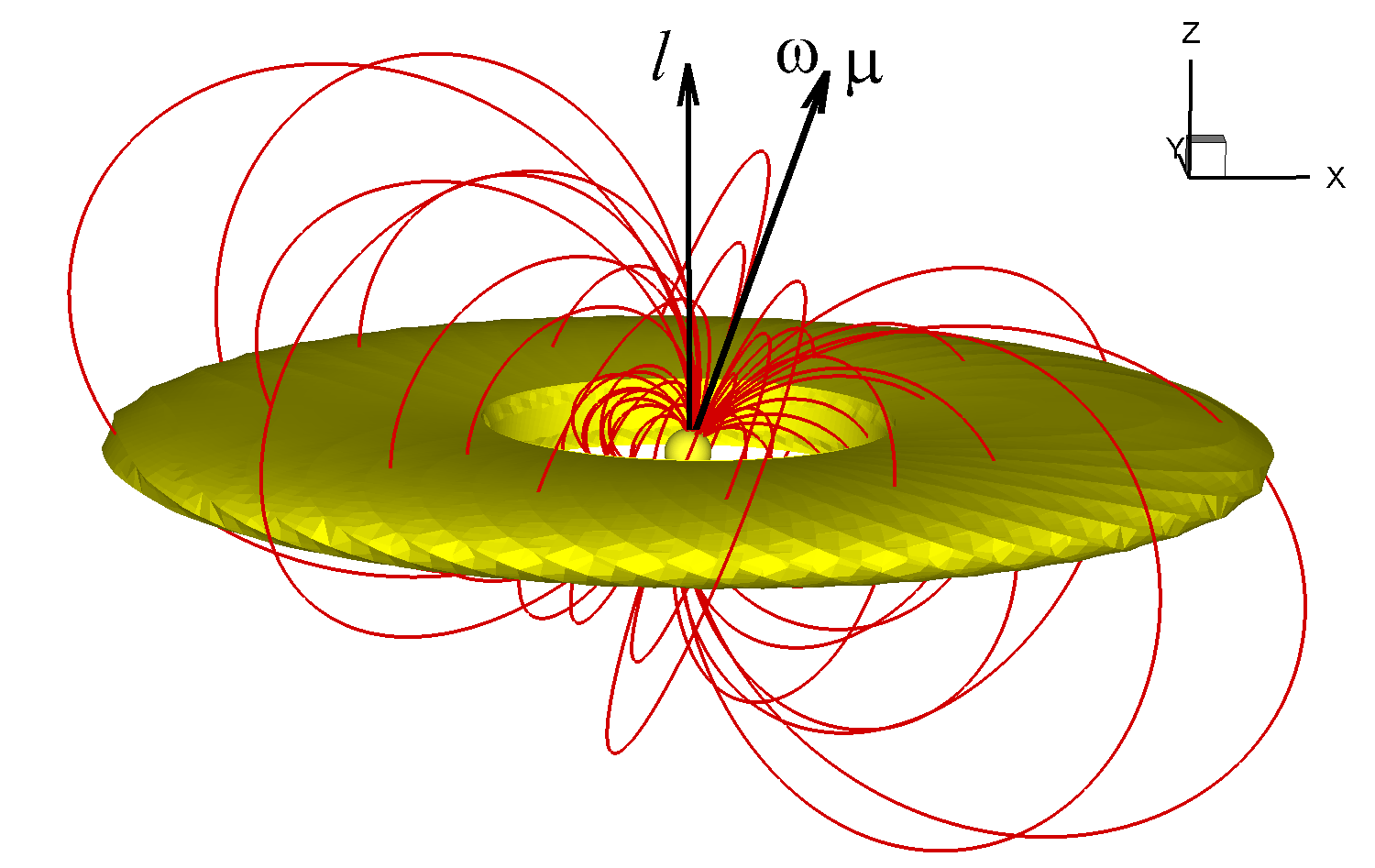}
\caption{3D view of the disc at $t=0$ in Model A.
The color background shows one of
the density levels ($\rho=1.7$ in dimensionless units). Lines are selected field lines. Arrows
show directions of the 
disc's angular momentum vector, $\bl$, the 
star's angular velocity vector,
$\bomega$, and the stellar dipole moment, ${\bmu}$. In this and other 3D plots we use the coordinate system $xyz$, where the $z-$axis
is along the initial direction of $\bl$ (see Fig. \ref{sketch}).}\label{3d-initial}
\end{figure}

Interaction of the inner disc with the tilted magnetosphere leads to 
bending torques in the disc, which result in a warp (bending wave)
in the inner disc
(e.g., \citealt{BouvierEtAl1999,TerquemPapaloizou2000,RomanovaEtAl2013}). If the rotation axis of the star is aligned with that of the disc, the warp rotates with the period of the star, 
and on average, the bending torque on the inner disc is zero (e.g., \citealt{Lai1999}).
However, if the rotational axis of the star is tilted about the rotational axis of the disc, then 
the time-averaged torque on the inner disc is not zero, and the inner parts of the disc may be warped systematically  
 (e.g., \citealt{Aly1980,LipunovShakura1980,Lai1999}).
The magnetic torque also may
drive the tilted inner disc into retrograde precession (opposite
to the rotation of the disc) around the rotational axis of the star.
Under some conditions
the combined effects of differential precession and viscosity tend to drive the inner disc toward an aligned state,
 where the disc plane lies in the rotational equator of the star 
 (the magnetic Bardeen-Petterson effect, \citealt{Lai1999}).  
Thus, the magnetic warping torque and the magnetic Bardeen-Petterson effect have an opposite consequence in the inner disc orientation; which effect wins depends on the dissipative properties of the inner disc and other parameters (such as the tilt of the magnetosphere). 

In theoretical studies, the configuration of the magnetosphere interacting 
 with the disc
was presented in the analytical form and was not fully time-dependent 
(e.g., \citealt{Aly1980,LipunovShakura1980,Lai1999}). 
Here, we show results of the global 3D MHD time-dependent numerical simulations of this problem, 
where the configuration of the magnetic field varies in time and depends on the relative motion of the rotating star and
the disc. 
 
Earlier, we performed global 3D MHD simulations of accretion onto a star with a tilted dipole magnetosphere, where the rotational axis of the star was aligned 
with the rotational axis of the disc (\citealt{RomanovaEtAl2003,RomanovaEtAl2004,RomanovaEtAl2013,RomanovaOwocki2015}; see also \citealt{ZhilkinBisikalo2010}). In our new simulations, we 
study numerically accretion onto stars where both the rotational and magnetic axes are tilted.

We performed simulations at a variety of different parameters and observed that a significant part of the disc becomes tilted.
 However, in some models
the disc normal tends to be aligned with the rotational axis of the star (aligned discs), while in other models it
 is systematically tilted.
 Comparisons of models show that
  an important parameter determining the final tilt is the position of the inner disc relative to the dipole magnetosphere. Higher tilts were observed in models where the disc is closer to the star and stronger magnetic field threads the disc. Another  parameter is the rotation of the star: higher tilts are observed in stars with faster rotation.

 In Sec. \ref{sec:theory} we review the theory.
In Sec. \ref{sec:numerical model} we describe our numerical model.  In Sec. \ref{sec:ABC}, \ref{sec:DEF} and \ref{sec:alignment}  we
show results of simulations. Discussion and conclusions are given in Sec.
\ref{sec:discussion}.

\section{Overview of the theory}
 \label{sec:theory}

Below, we briefly review the theory 
 following the approach
of Lai (1999) and \citet{FoucartLai2011} (see also \citealt{Aly1980,LipunovShakura1980,LaiEtAl2011}).

\subsection{Warping instability}
\label{sec:warping}

We consider a star of mass $M_\star$ and radius $R_\star$ which rotates with an angular velocity $\bomega=\omega \hat{\bomega}$, where  $\hat{\bomega}$
is the unit vector. 
 The rotational axis of the star $\hat{\bomega}$ is tilted about the disc's angular momentum vector 
 $\bl$ by an angle $\beta$. We suggest that a star has a dipole magnetic field and place the magnetic moment
$\bmu$ at an angle $\theta$ relative to $\bomega$. Vector $\bmu$ rotates about $\bomega$ with angular velocity of the star, $\omega$  (see sketch in Fig. \ref{sketch}).

Matter of the disc accreting with the rate $\dot M$ is stopped by the magnetosphere of the star at the magnetospheric radius (e.g., \citealt{PringleRees1972,GhoshLamb1978}):
\be r_m=k
\left({\mu^4\over GM_\star\dot M^2}\right)^{1/7}, \label{alfven}
\ee
where $k\approx 0.5$  (e.g., \citealt{LongEtAl2005,BessolazEtAl2008}). 

The tilted magnetosphere interacts with the inner parts of the accretion disc. Such interaction may lead  to warping and precession of the disc.
For analysis of the disc warping, we use the coordinate system $xyz$, with the $z-$axis initially directed along the disc normal $\hat{\bl}$. We use the  variable $\hat{\bl}$ to indicate the initial position of the disc normal. 
We suggest that the direction of the disc normal may change in time and we use the variable $\hat{\bl}_n$ for the disc normal of  warping disc. We also use the variable  $\beta_n$ for changing angle between the disc normal  $\hat{\bl}_n$ and $\hat{\bomega}$.

  The vertical (perpendicular to the disc) magnetic
field produced by the stellar dipole is given by
 \be
B_{z}=-{\mu\over
r^3}\left(\cos\theta\cos\beta_n-\sin\theta\sin\beta_n\sin\omega
t\right).
\label{eq:Bz} 
 \ee 
We assume that the static field component,
$B_{z}^{s}=-(\mu/r^3)\cos\theta\cos\beta_n$, penetrates the disc
in an ``interaction zone'', between $r=r_m$ and $r_{\rm int}$. This
field is twisted by the differential rotation between the star and
the disc.
The toroidal field at the disc increases in time until it becomes
comparable to $|B_{z}^{s}|$, at which point the magnetic field lines inflate (e.g., \citealt{LovelaceEtAl1995}).
Here, we suggest that the field associated with the twist of the magnetic field lines is 
equal above and below the disc, with the only difference in the direction of the field:
  $B_{\phi}'(r)=\mp\zeta B_{z}^{s}(r)$, where parameter $\zeta\sim 1$. There is also a toroidal component of the dipole field, which has the same sign above and below the disc: 
$B_{\phi}^{\mu}=-(\mu/r^3)(\hat\bmu\cdot\hat\phi)$, where
$\hat\phi$ is the unit vector in the azimuthal direction around 
the disc. Thus there is a vertical magnetic force on the disc which is the difference in the magnetic pressure between the lower and upper sides of the disc:
\be
F_{z}(r)={1\over 8\pi}\left[(B_{\phi}^{\mu}+\zeta
B_{z}^{s})^2- (B_{\phi}^{\mu}-\zeta B_{z}^{s})^2\right]
={\zeta\over 2\pi}B_{\phi}^{\mu} B_{z}^{s}. 
\label{eq:Bz}
\ee 
There is a torque acting on the disc, which leads to warping instability. 
The torque per unit area on the disc can be calculated by 
averaging over the azimuthal angle in the disc and the stellar
rotation period,  
\be \bN_w(r)=-{\zeta\mu^2\over 4\pi
r^5}\cos\beta_n\cos^2\!\theta
\,{\hat{\bl}_n}\times(\hat\bomega\times{\hat{\bl}_n}). \label{eq:N_w}\ee
For $\zeta>0$, the effect of
this torque
is to push the
local disc axis $\hat{\bl}_n$ away from $\hat\bomega$ toward the
``perpendicular'' state. 
The characteristic warping rate is 
\be \Gamma_w(r)=\frac{\zeta\mu^2}{4\pi
r^7\Omega(r)\Sigma(r)}\cos^2\theta, \label{eq:Gamma_w} \ee where
$\Sigma(r)$ is the surface mass density of the disc and  $\Omega(r)$ is the angular velocity of the disc. 

The disc is expected to be warped (or tilted) up to the distance where the time scale of warping $t_w=\Gamma_w^{-1}$ becomes comparable with the viscous time scale, $t_v=r^2/\nu_2$, where $\nu_2$ is the $rz-$component of viscosity (perpendicular to the disc).  
The warping radius is of the order of the magnetospheric radius $r_m$ (see  Eq. 4.12 in \citealt{Lai1999}).

\begin{table*}
\centering
\begin{tabular}{l||llllllllllllll}
\\ Model   &   $\beta$     & $\theta$  & $\mu'$  & $M_d$ & $r_{\rm in}$ & $r_c$ & $P_\star$ & $\bar{r}_m$ & ${\bar r}_m/r_c$ & $r_t$ & $r_t/{\bar r}_m$ & $\tau_{\rm sim}/P_0$&$\tau_{\rm sim}/P_\star$ & $\beta_t$  \\
\hline
$\bf A$    &  $20^\circ$   & $2^\circ$   &  1   &  $M_{\rm d0}$    &      8.6    & 14.3& 11.2   & 3.4 &0.24& $ 24.3$ &$ 7.3$& 150 &13.4& $5^\circ-10^\circ$   \\
\hline
$\bf B$    &  $20^\circ$  &  $20^\circ$  & 1   &  $0.3M_{\rm d0}$ &      8.6    &14.3 & 11.2  & 4.0 &0.28& $ 21.8$&$ 5.4$& 100 &8.9& $5^\circ-10^\circ$  \\
\hline
$\bf A1$  &  $20^\circ$  &  $2^\circ$    & 1  &  $0.3M_{\rm d0}$  &      8.6    &14.3 & 11.2  & 4.0 &0.28 &$ 21.4$& $ 5.3$& 120 &10.7& $5^\circ-10^\circ$   \\
\hline
$\bf B1$  &  $20^\circ$  &  $20^\circ$  & 1   &  $M_{\rm d0}$     &      8.6    &14.3  & 11.2  & 3.4 &0.24 &$ 21.1$ & $ 6.2 $& 180 &16.1& $5^\circ-10^\circ$  \\
\hline
$\bf C$   &  $15^\circ$   &  $2^\circ$   & 1   &  $M_{\rm d0}$     &      8.6     &8.6   & 5.2    & 3.4 &0.39&$ 20.9$ & $ 6.2$& 120 &23.1& $15^\circ-20^\circ$ \\
\hline
$\bf D$    &  $15^\circ$  &  $2^\circ$  & 0.5  &  $M_{\rm d0}$    &      5.7     &8.6   & 5.2   & 2.9 &0.34&$ 19.0$ & $6.5$ & 70 &13.5& $30^\circ-40^\circ$      \\
\hline
$\bf E$    &  $15^\circ$  &  $2^\circ$  & 0.5  &  $M_{\rm d0}$    &      5.7     &5.1   & 2.4   & 2.9 &0.57&$19.4$ & $ 6.5$& 60 &25.0& $30^\circ-40^\circ$     \\
\hline
$\bf F$    &  $15^\circ$  &  $15^\circ$ & 0.3  & $M_{\rm d0}$     &      5.7     &5.1   & 2.4    & 2.1 &0.41&$ 17.7 $ &$ 8.6$& 60 &25.0& $30^\circ-40^\circ$     \\
\hline
\hline
\end{tabular}
\caption{Representative simulation models. From left to right: tilt angles of the rotational and magnetic axes of the star
relative to the disc normal, $\beta$ and $\theta$, respectively; the magnetic moment of the star, $\mu'$; mass of the disc, $M_d$; initial radius of the inner disc, $r_{\rm in}$; corotation radius, $r_c$; period of the star, $P_\star$; time-averaged magnetospheric radius, ${\bar r}_m$; the ratio  ${\bar r}_m/r_c$; radius of the tilted disc, $r_t$;  the ratio $r_t/{\bar r}_m$; $\tau_{\rm sim}/P_0-$duration of simulation runs (in periods $P_0$ of Keplerian rotation at $r=1/0.35$); $\tau_{\rm sim}/P_\star-$duration of simulation runs in periods of the star; the tilt angle, $\beta_t$.}
\label{tab:models}
\end{table*} 

\subsection{Precession of the disc}
\label{sec:precession}

There is also a precessional
torque on the disc. 
The torque arises from the
dielectric property of the disc. If the disc does not allow the
vertical stellar field 
to penetrate, an azimuthal
screening current $K_{\phi}$ is induced in the disc. 
It interacts with the radial magnetic field $B_{r}$ from
the stellar dipole and produces a vertical force. After azimuthal
averaging and averaging over the stellar rotation, we obtain the
torque per unit area:
\begin{equation}
 \bN_p(r)=\frac{\mu^2}{\pi^2 r^5 D(r)}
\cos\beta_n\Omega_p(r)\,\hat\bomega \times\hat{\bl}_n ,
\label{eq:N_p}
\end{equation}
where $D(r)$ is
a function of $r/r_m$ and $h(r)/r_m$, where $h(r)-$is the half-thickness of the disc  (see Eq. 2.4 from \citealt{Lai1999}).
The torque $\bN_p(r)$ 
pushes the disc to precess
around the rotational axis of the star. The precession angular frequency is ${\bf\Omega}_{\rm
prec}(r)= -\Omega_p(r)\cos\beta\,\hat\bomega$, where
\begin{equation}
 \Omega_p(r)=\frac{\mu^2}{\pi^2 r^7\Omega(r)\Sigma(r) D(r)} F(\theta) ,
\label{eq:Omega_p}
\end{equation}
where $F(\theta)=2f\cos^2\theta-\sin^2\theta$ . 
Parameter $f=1$, if the stellar vertical component is entirely screened from the disc, and  $f=0$, if only the time-varying component is screened out.

\subsection{Magnetic Bardeen-Petterson effect }
\label{sec:BP}

The combination of viscous and precession torques may lead to the gradual alignment of the inner disc with the equatorial plane 
of the star. 
This phenomenon has been extensively studied in cases of  non-magnetic stars where a disc undergoes the  Lense-Thirring precession around a rotating compact object
(e.g., \citealt{BardeenPetterson1975, PapaloizouPringle1983,KumarPringle1985,KumarPringle1992,Pringle1992,
ScheuerFeiler1996,IvanovIllarionov1997,Ogilvie1999,LubowEtAl2002,FragileEtAl2007}). 

In magnetized stars both, the tilt of the disc and its precession are driven by the magnetic force. One can derive the
magnetic Bardeen-Petterson radius in analogy with the approach used for relativistic stars \citep{Lai1999}.
Setting the precession time scale ${\Omega_p(r)}^{-1}$ equal to the viscous time scale, $r^2/\nu_2$ (where $\nu_2$ is viscosity coefficient in the direction perpendicular to the disc ), one obtains the magnetic
Bardeen-Petterson radius\footnote{\citet{KumarPringle1985} provided a more precise approach to the problem.  However, in application to magnetized stars we follow an approximate approach of \citet{Lai1999}.}:
Radius  $R_{\rm MBP}$ is of the same
order as  the warping radius $r_w$
inside which 
the disc tilt grows  (see Eq. 4.14 in \citealt{Lai1999}). 
Which effect dominates depends on the dissipative properties of the inner disc (see
also \citealt{FoucartLai2011}).

\section{Numerical model}
\label{sec:numerical model}
  
We perform global 3D MHD simulations of matter accretion onto a magnetized star with tilted magnetic and rotational axes.
We use the earlier developed code   \citep{KoldobaEtAl2002} which is modified to incorporate the tilt of the rotational axis. Below, we briefly describe our model. 

\subsection{Initial and boundary conditions.} 

\noindent{\bf Initial conditions.} We place the accretion disc in the $xy$ plane such that its normal vector $\hat{\bl}$ is tilted about the rotational axis of the star by an angle $\beta$ (see Fig. \ref{sketch}).

The disc is cold and
dense, while the corona is hot and rarefied, and at the reference
point (the inner edge of the disc in the disc plane at $t=0$), the disc is 100 times denser than corona, while the temperature of the disc
is 100 times lower.

Initially, the disc and corona are in 
the rotational hydrodynamic equilibrium (see, e.g.,  \citealt{RomanovaEtAl2002}).
The initial conditions are derived from the balance of the gravitational, centrifugal, and pressure gradient forces. 
Initially, we rotate both the disc and corona 
with Keplerian velocity $v_K(r)$. This condition helps to eliminate the effects of the initial discontinuity of the magnetic field lines at the disc-corona boundary. \footnote{In the opposite case 
strong magnetic braking of the disc and rapid accretion have been observed.}  The corresponding distributions of density and pressure were derived analytically (see Eqs. 5-10 in \citealt{RomanovaEtAl2002}). The top left panel of Fig. \ref{2d-initial-8} shows a typical density distribution in the disc.\footnote{Note that this density distribution does not correspond to the viscous equilibrium, and we usually observe that the density in the disc is slowly redistributed on the viscous time scale.} 

In all models, we consider discs with the same initial density and temperature at the fiducial point (at the inner disc). To vary the mass of the disc we change the initial disk thickness  $h(r)/r$.   Soon after the beginning of simulations, the thin disc expanded and became thicker, because we took the same initial sound speed in all models (corresponding to $h(r)/r\approx 0.1-0.15$). However, discs with smaller initial values of $h(r)/r$ have $\sim 3$ times smaller mass. 
The top and bottom leftmost panels of Fig. \ref{2d-initial-8} show the initial configurations of the disc and magnetosphere for the more massive (top panel) and less massive (bottom panel) discs. Fig. \ref{3d-initial} shows a 3D view of the initial configuration in one of the models.

The size of the simulation region is $r_{\rm out}\approx 34 R_\star$. Initially, we place the inner disc at distances    
$r_{\rm in}\approx 8.6 R_\star$  or $5.7R_\star$  which are larger than expected magnetospheric radii $r_m$. This helps to start simulation smoothly. Later, the disc moves inward and settles at the magnetospheric radius.  

\smallskip

\noindent\textbf{Boundary conditions.}
  At the inner boundary (stellar surface) and the outer
boundary, most of the variables $F_i$ have free
boundary conditions, ${\partial F_i}/{\partial r}=0$.
We fix  the normal component of the field, $B_n$ to support the frozen-in condition.

\begin{figure*}
\centering
\includegraphics[width=16cm]{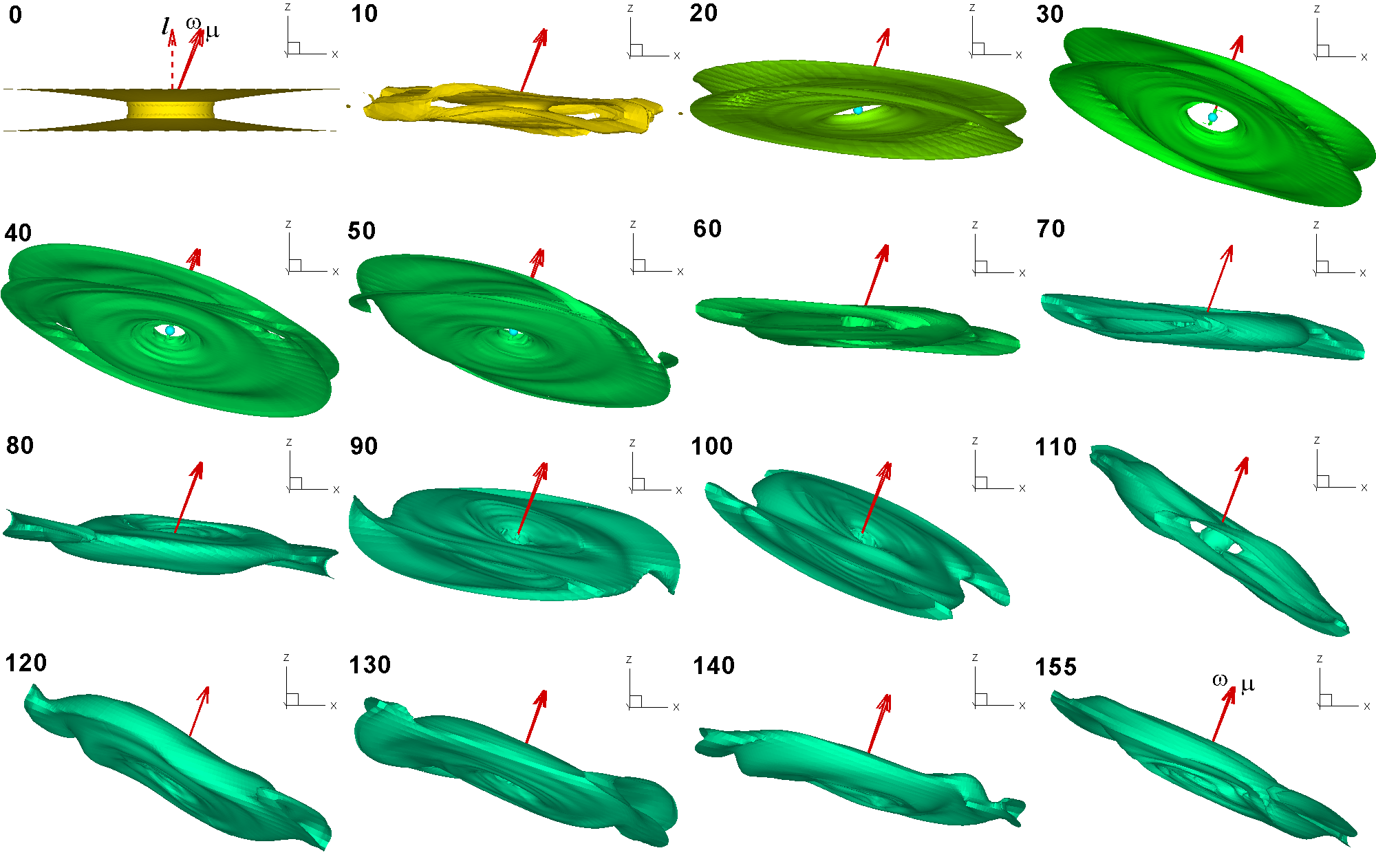}
\caption{3D views of the disc in Model A at different moments of time. 
The color background shows density levels which vary from 
$\rho=0.75-1.5$ in the top row of panels, to $\rho=0.23-0.54$ in three bottom rows of panels.} \label{3d-t2-bet20-16}
\end{figure*}

\begin{figure*}
\centering
\includegraphics[width=12cm]{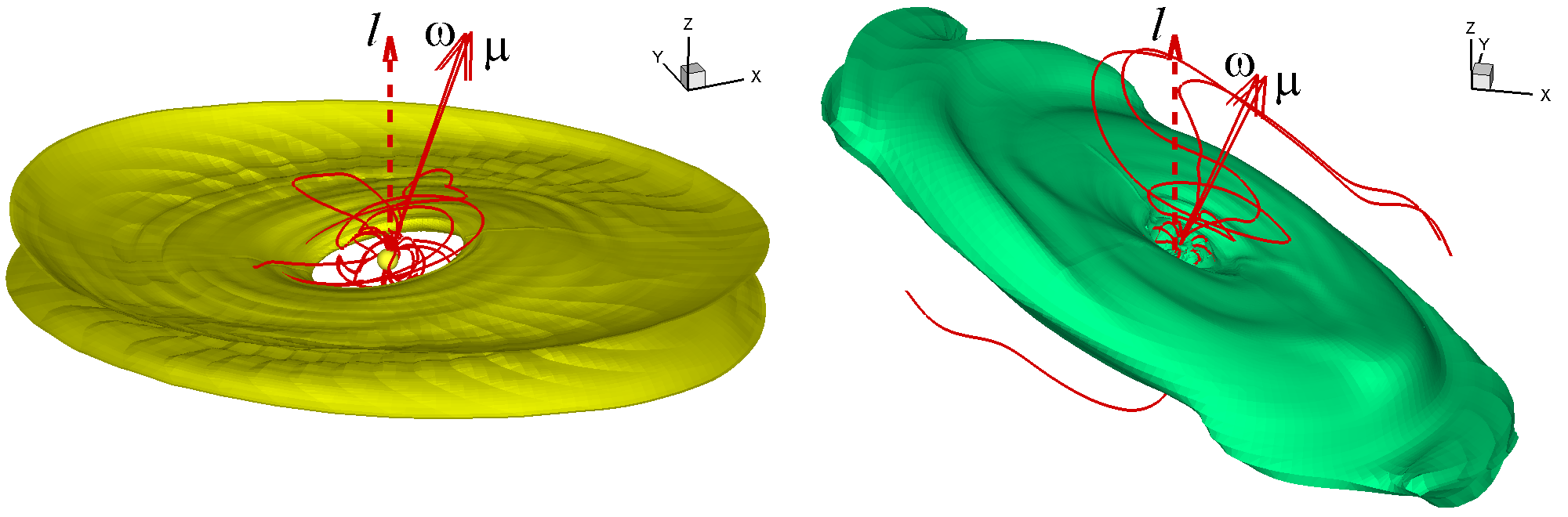}
\caption{3D views of the disc in Model
A  at moments $t=18$
(left) and $t=120$ (right), respectively. The color background
shows the density levels. Lines are sample magnetic field lines.}
\label{chi20-d1-c3-5-b001}
\end{figure*}

\subsection{Code description and dimensionalization}
 
\noindent{\bf The code.} We solve the 3D MHD equations with a Godunov-type code in a reference frame rotating with the star, using the ``cubed sphere" grid \citep{KoldobaEtAl2002}. We use the 8-waves Roe-type approximate Riemann’s solver analogous to that described by  \citet{RuyJones1995}. We split the magnetic field to that of the star  and  induced by currents in the disc and corona.

In this work, we use the entropy balance equation instead of the full energy equation because we do not expect shocks inside the simulation region. \footnote{Shocks are expected at the stellar surface. However, this problem has been studied separately, on different spatial scales (e.g., \citealt{KoldobaEtAl2008}).}

\smallskip

\noindent{\bf Viscosity.} The viscosity term is incorporated 
into the momentum equation
with the $\alpha-$prescription
for the viscosity coefficient  $\nu\sim \alpha p$, where $p$ is pressure in the disc \citep{ShakuraSunyaev1973}. 
  The viscosity is nonzero only inside the disc, above a  threshold
density ($\rho_v = 0.1\rho_d$, where $\rho_d$ is the density in the disc). We use 
 $\alpha=0.02$ in all simulation runs.  
In reality, the
disc is expected to be turbulent, where turbulence can be driven by the  magneto-rotational instability (MRI, e.g., \citealt{BalbusHawley1991}).
\footnote{Axisymmetric and 3D simulations of accretion from turbulent MRI-driven disc 
have shown many similarities in properties of magnetospheric accretion compared with $\alpha-$discs
 \citep{RomanovaEtAl2011,RomanovaEtAl2012}.} However, these simulations are time-consuming.

\begin{figure*}
\centering
\includegraphics[width=16.0cm]{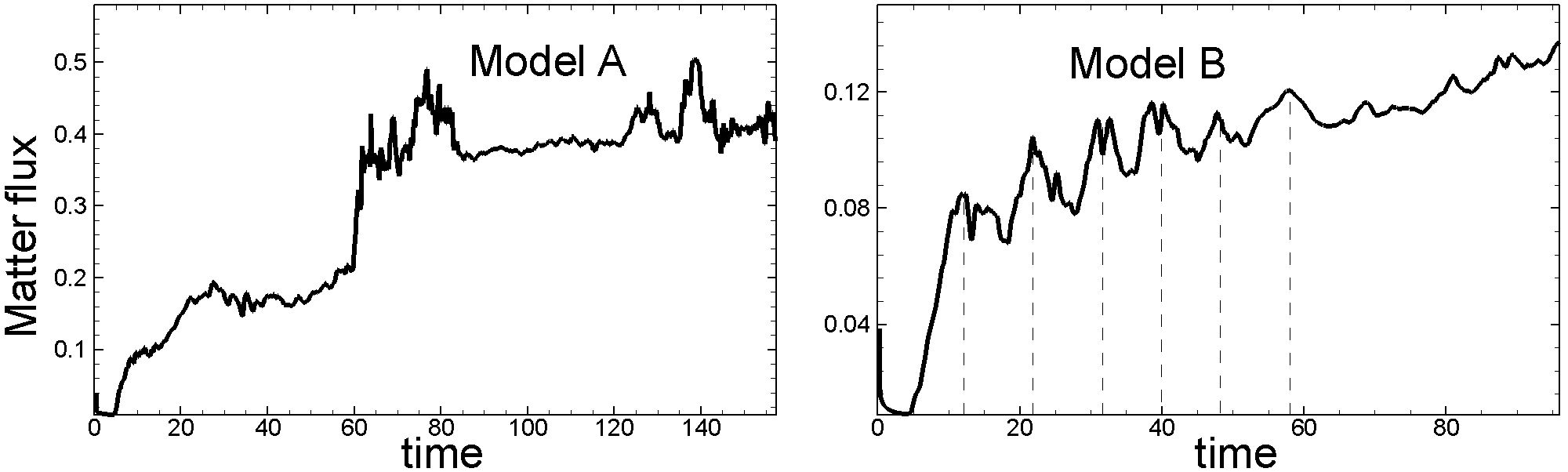}
\caption{Temporal variation of matter flux $\dot M$  (in dimensionless units, see Tab. \ref{tab:refval})  at the surface of the star in models A and B (left and right panels, respectively). Dashed vertical lines in right-hand panel show peaks of  $\dot M$ corresponding to moments of high tilt of the magnetic axis relative to the disc normal.} \label{pm-all-2}
\end{figure*}

\smallskip

\noindent\textbf{The grid} consists of $N_r$ spheres. Each sphere
represents an inflated cube with six sides. 
Each side has a $N\times N$ curvilinear grid, which represents a
projection of the Cartesian grid onto the sphere. The whole grid
consists of $6\times N_r\times N^2$ cells. We use the grid with 
$N_r=140$ and  $N = 61$.
The MPI-parallelized code uses 28 layers in the radial direction and 6 layers for six sides of the inflated cube, with  168 layers total.

\smallskip

\begin{table}
\begin{tabular}{l@{\extracolsep{0.2em}}l@{}lll}

\hline
&                                                   & CTTSs       & White dwarfs          & Neutron stars           \\
\hline
\multicolumn{2}{l}{$M_\star (M_\odot)$}                    & 0.8         & 1                     & 1.4                     \\
\multicolumn{2}{l}{$R_\star$}                             & $2R_\odot$  & 5000 km               & 10 km                   \\
\multicolumn{2}{l}{$R_0$ (cm)}                      & $4\e{11}$   & $1.4\e9$              & $2.9\e6$                \\
\multicolumn{2}{l}{$P_0$}                                 & 1.8 days       & {29 s}               & {2.2 ms}                  \\                     
 \multicolumn{2}{l}{$B_{\star0}$ (G)}               & $10^3$      & $10^6$                & $10^9$                  \\
\multicolumn{2}{l}{$B_0$ (G)}                       & 43          & $4.3\e4 $              & $4.3\e7 $  \\
\multicolumn{2}{l}{$\rho_0$ (g cm$^{-3}$)}          & $7\e{-12}$  & $2\e{-8}$             & $2.8\e{-5}$  \\
\multicolumn{2}{l} {{$\dot M_0$} ($M_\odot$yr$^{-1}$)}   &   $2.8\e{-7}$  &  $1.9\e{-7}$   &   $2.9\e{-8}$  \\     
\hline                     
\end{tabular}
\caption{Sample reference values for three types of stars. }
\label{tab:refval}
\end{table}

\noindent\textbf{Dimensionalization.}
Equations are solved in dimensionless form.  
The dimensionless variables are determined as  $\tilde F=F/F_0$,  where  $F$ are dimensional variables, while  $F_0$ are their reference values. 
 The reference value of distance $R_0$ is chosen such that the star has radius $R_\star = 0.35R_0$. The reference velocity is the Keplerian velocity at $R_0$, $v_0 = (GM_\star/R_0)^{1/2}$.
The reference time is $t_0 = R_0/v_0$.   
The magnetic moment of the star: $\mu_\star=\mu' B_{\star 0} R_\star^3$, where
 $B_{\star_0}$
is the reference surface magnetic field of the star at the magnetic equator and $\mu'$ is dimensionless magnetic moment,  which helps to vary the magnetic field of the star: 
$B_\star=\mu' B_{\star_0}$. 
 The reference magnetic field, $B_0$, is the value of the magnetic field at $r=R_0$: $B_0=B_{\star_0}(R_\star/R_0)^3$. 
The reference density and pressure are $\rho_0 = B_0^2/v_0^2$ and $p_0 = \rho_0 v_0^2$, respectively. 

We take into account that $R_0=R_\star/0.35\approx 2.86 R_\star$ and for convenience show distances in radii of the star. Also, we show time in periods of rotation at this radius,   $P_0 = 2\pi R_0/v_0$.
 Below, we use dimensionless variables but drop tildes. The results of simulations can be applied to stars of different types. 
Table \ref{tab:refval} shows sample reference values for different types of stars.

\begin{figure*}
\centering
\includegraphics[width=16cm]{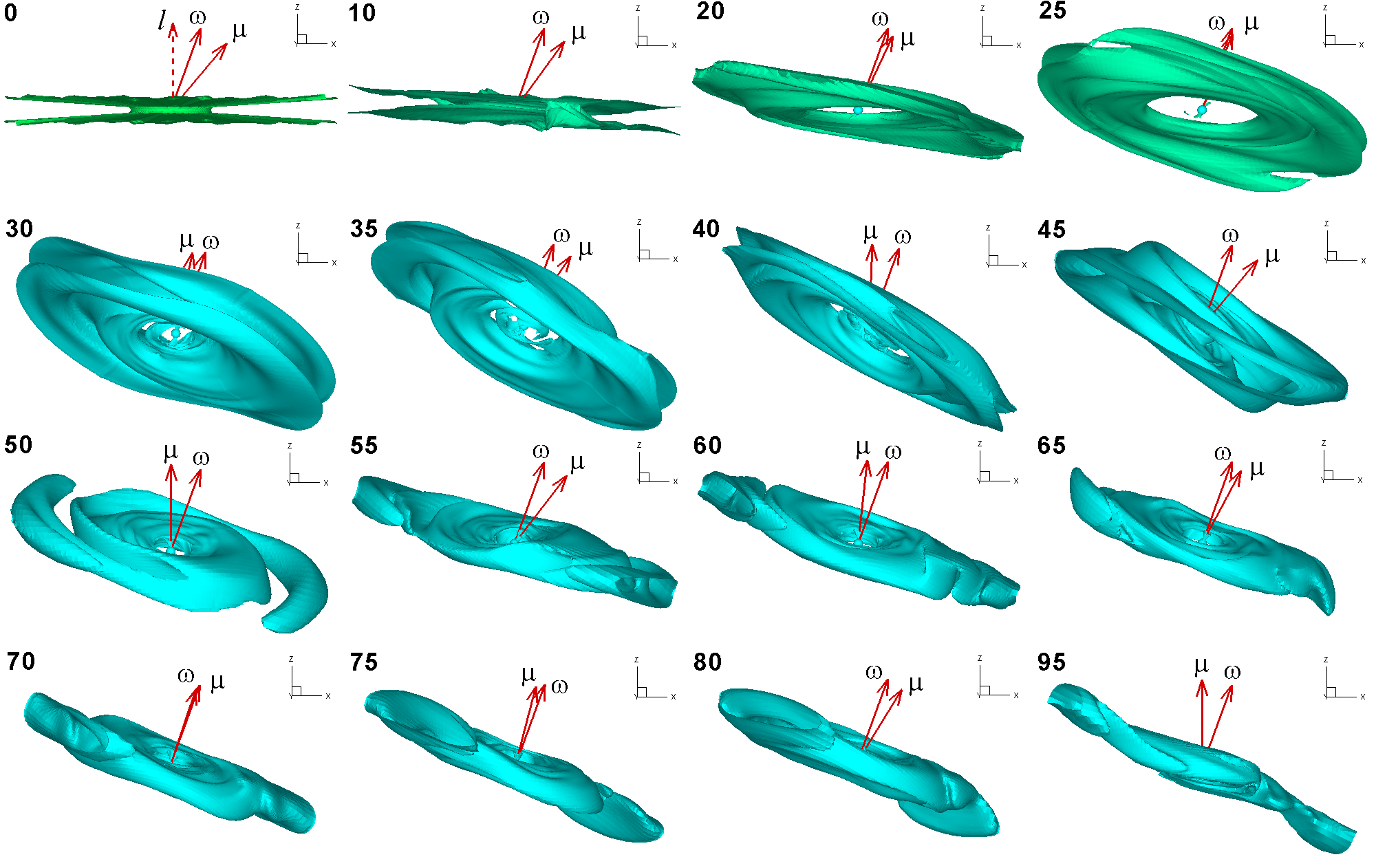}
\caption{3D views of the disc at different moments of time in Model B. 
The color background shows density levels which vary from 
$\rho=0.3-1.5$ in the top row of panels, to $\rho=0.15-0.2$ in three bottom rows of panels. 
} \label{3d-t20-bet20-16}
\end{figure*}

\subsection{Set of models}

We performed simulations at a variety of different parameters:
 different initial inclination angles of the rotational axis:  $\beta=15^\circ$ and $20^\circ$;
small and relatively large tilt angles of the dipole: $\theta=2^\circ$,  $15^\circ$ and  $20^\circ$;\footnote{We took a small  angle, $\theta=2^\circ$ because in the case of  $\theta=0^\circ$, a stronger switch-on wave is observed and a more gradual spin-up of the star is required at the beginning of the simulation.}
different values of the dipole moment: $\mu'=1, 0.5, 0.3$;
different values of the rotational period of the star, which varied from  $P_\star=11.2$ to $P_\star=2.4$ \footnote{In the code we determine the period of the star using the corotation radius
 $r_c$, which is the radius where the
angular velocity of the disc matches the angular velocity of the star, $\Omega(r_c)=\omega$, $r_c=(GM_\star/\omega^2)^{1/3}$. }. We also varied the initial position of the inner disc, $r_{\rm d_in}$. Table \ref{tab:models} shows parameters of models.

Simulations show that in all models the inner disc was warped, then tilted, and became approximately flat. However, in some models, the
normal to the tilted disc, ${\bl}_t$ tends to align with the rotational axis of the star, and typical tilt angles are small, $\beta_t\approx 5^\circ-10^\circ$ (we call them aligned discs).
In other models, the disc normal is tilted at a larger angle, $\beta_t\approx 30^\circ-40^\circ$ (we call them tilted discs).
 Comparisons of results at different sets of parameters
showed that one of the main parameters is the initial position of the inner disc, $r_{\rm in}$.
When we place the inner disc at larger distances,
$r_{\rm in}=8.6$, we obtain only slightly tilted (aligned) discs.
In the opposite case,
 $r_{\rm in}=5.7$, \footnote{Note that these radii (measured in stellar radii for convenience) result from ratios $2/0.35$ and $3/0.35$ and correspond to $r_{\rm in}=2$ and $3$ (in units of $R_0$).} we obtain discs with larger tilts. Another important parameter is the corotation radius, $r_c$: at smaller values of this parameter (faster rotating stars) we obtain
 discs with larger tilts.
 Below, in Sec. \ref{sec:ABC} and \ref{sec:DEF}, we consider two groups of models corresponding to two values of $r_{\rm in}$, and different values of $r_c$.

\section{Models of aligned or slightly tilted discs (A, B, A1, B1, C)}
\label{sec:ABC}

In several models, we placed the inner radius of the disc at a relatively large distance from the star, $r_{\rm in}=8.6$.
We considered two main models, A and B.  In both models,
 the rotational axis of the star is tilted by $\beta=20^\circ$, while the tilt angles of the magnetosphere are different: $\theta=2^\circ$ in Model A and $\theta=20^\circ$ in Model B.
 In Model A, we took a disc of higher mass, while in Model B
 the disc has three times lower mass.  In these models, we took the corotation radius $r_c=14.3$ which corresponds to a slow rotation of the star.
 We also considered three supplement models. Models A1 and B1 are identical to models A and B, but the disc mass is $\sim 3$ times lower/ higher, respectively. Model C is identical to Model A, but a star rotates more rapidly: $r_c=8.6$.

\subsection{Accretion onto a star with a tilted rotational axis: $\beta=20^\circ$, $\theta=2^\circ$ (Model A)}
\label{sec:A}

In this model, we test the main new feature - how the inner disc evolves in the case when the rotational axis of the star is tilted about the rotational axis of the disc, while the magnetic axis is almost aligned.

We observed that the disc initially moved towards the star and was stopped 
 by the magnetosphere at the distance $r_m$ where matter pressure in the disc equals the magnetic pressure of the magnetosphere \citep{PringleRees1972}, that is where the modified plasma parameter $\beta_1=8\pi (p+\rho v_\phi^2)/B^2=1$. At this distance, matter started flowing to the star in funnel streams (or in unstable tongues, e.g.,  \citealt{KulkarniRomanova2008}).  We used the condition $\beta_1=1$ in the equatorial plane to find the magnetospheric radius. This radius slightly varies in time due to variability in accretion rate.  The time-averaged value is $\bar{r}_m\approx 3.4$.
The top right panel of Fig. \ref{2d-initial-8} shows the close view of matter flow near the star and $\beta_1=1$ line. 
Top middle panels of the same figure show $x'z'-$slices of density distribution and poloidal field lines at  $t=15$ and $20$. One can see that the field lines inflate and become non-dipolar in most of the simulations region, excluding the inner parts of the disc, where the modified dipole can be seen (see the right-hand panel of the same figure).

\begin{figure*}
\centering
\includegraphics[width=16.0cm]{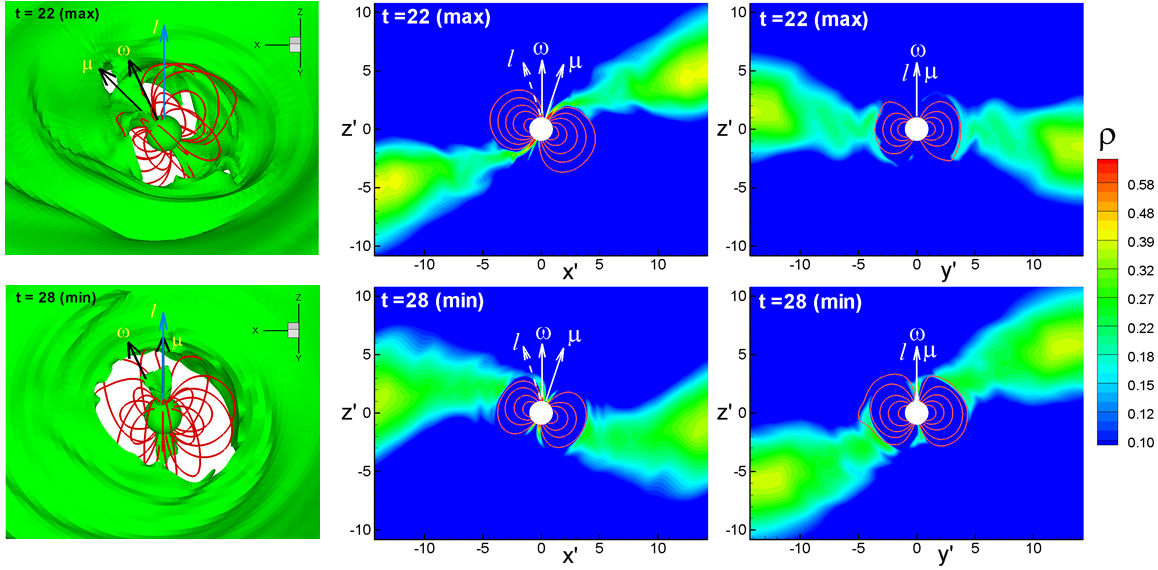}
\caption{\textit{Left panels:} 3D views of the inner disc during the episodes of the
local maximum (top) and minimum (bottom) accretion rates in 
Model  B at moments $t=22$ and $t=28$, respectively.
The color background shows one of density levels.  Lines are
sample field lines.  \textit{Middle and right panels:} $x'z'$ and $y'z'$ slices of density distribution and sample field lines
corresponding to the same moments in time. }\label{2d-3d-min-max}
\end{figure*}

The magnetic force  and warping torque rapidly decrease with the distance from the star   
 (see Eq. \ref{eq:N_w} for torque), and therefore they act mainly in the proximity of the disc-magnetosphere boundary. 
However, we see that  a significant part of the disc becomes tilted. 
We suggest that information about the inner warp propagates to larger distances in the form of bending waves. 
According to \citet{PapaloizouPringle1983} and \citet{PapaloizouLin1995}, the disc may be either in the diffusive regime (if $h(r)/r<\alpha_2$), or in wave regime  (if $h(r)/r>\alpha_2$). 
In our simulations  
$\alpha_2\approx0.02$, 
the ratio $h(r)/r \approx 0.1-0.15$,  $\alpha_2 < h(r)/r$, and  the disc 
is in the wave regime. 
\footnote{In our earlier 3D MHD simulations of waves generated by the tilted rotating dipole,
we observed that bending waves are generated by the warp and propagate to large distances \citep{RomanovaEtAl2013}. 
In these new simulations, we use similar code and expect that bending waves also propagate with little damping.}

Fig. \ref{3d-t2-bet20-16} shows 3D views of the disc at different times in Model A. 
We observed that  the inner parts of the disc were
warped, precessed, and tilted under the influence of the magnetic force,
as predicted by the theory (see Sec. \ref{sec:theory}). 
Initially, at $t=10-20$, the warp formed  in the inner parts of the disc. 
Later,  at $t>20$,  larger parts of the disc become warped and tilted. 
Subsequently, the significant part of the inner disc becomes tilted and almost flat.\footnote{Note that we use free boundary conditions at the outer boundary, which do not restrict the motion  along the outer boundary.} 
Fig. \ref{3d-t2-bet20-16} also shows that the disc can be split into two parts: the inner part, which is almost flat and has the same tilt, and the outer part with a different tilt (see, e.g., panels at $t=120, 130$ and $140$).
 We call the inner part the ``tilted disc''. Its time-averaged radius is $r_t\approx 24.3$ (in stellar radii) or 
 $r_t/{\bar r}_m\approx 7.3$ in magnetospheric radii (see Tab. \ref{tab:models}).

The disc slowly precesses about the rotational axis of the star. 
The rate of precession, $\Omega_p(r)$ (see Eq. \ref{eq:Omega_p}) depends on a number of factors, including the factor $f$ which characterizes the dielectric property of the disc. If only the time-varying component is screened, $f=0$, we obtain a factor $\sim\sin{\theta}\approx 0.035$. However, we observed comparable rates of precession in
models with $\theta=2^\circ$ and larger values of $\theta$.  We suggest that we have some intermediate situation, in which $0<f<1$.

\begin{figure*}
\centering
\includegraphics[width=16cm]{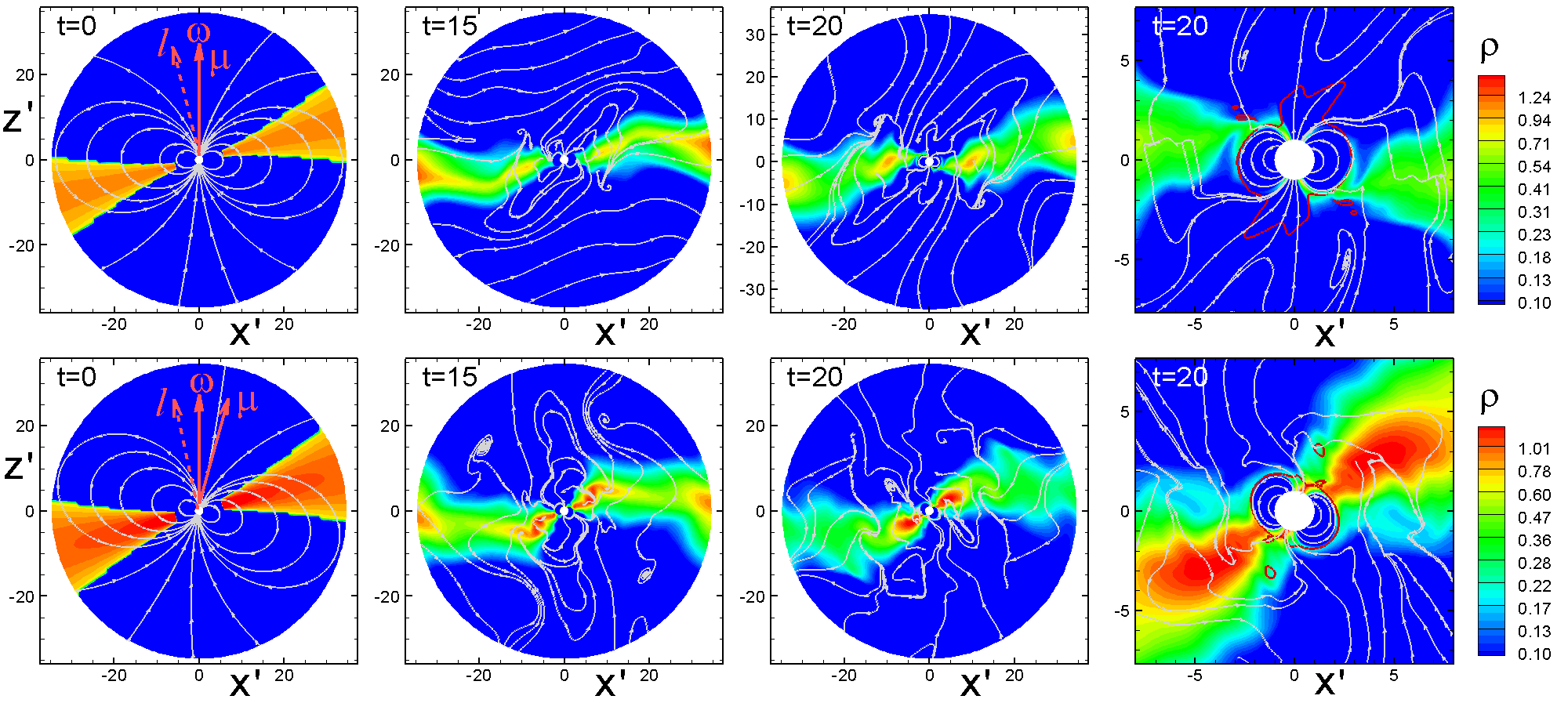}
\caption{$x'z'-$slices of density distribution in models D and F (top and bottom panels, respectively) at $t=0$,  $t=15$, and $t=20$. Top and bottom right panels show close view of matter flow near the magnetosphere at $t=20$. 
White lines show sample poloidal magnetic field lines. Red lines shows the $\beta_1=1$ line.}
\label{2d-d05-d03-8}
\end{figure*}

We observed that after a few periods of stellar rotation 
(approximately after $t=70$, see Fig.  \ref{3d-t2-bet20-16}),
 the disc starts tilting towards the equatorial plane of the 
star, so that the disc normal becomes almost parallel
to the angular velocity of the star, $\bomega$. There is still some tilt, but it is small, $\beta_t\lesssim 5^\circ-10^\circ$.
We discuss possible mechanisms of the disc alignment in Sec. \ref{sec:alignment}.
Fig. \ref{chi20-d1-c3-5-b001} shows typical initial and final states of the disc evolution.

The left-hand panel of Fig. \ref{pm-all-2} shows the accretion rate onto the star. 
 We observed 
 persistent accretion during 160 rotations (Keplerian periods at $r=1/0.35\approx 2.86$), which is approximately 14 periods of stellar rotation. Initially, the accretion rate increases due to the inward flow of the disc matter from the initial radius $r_i=8.6$ to the radius,  
where the disc is stopped by the magnetosphere, at $\bar{r}_m\approx 3.4$.  Later, at $20\lesssim t\lesssim 60$, matter accrets in two funnel streams, and accretion is quasi-stationary. 
 At $t\gtrsim 60$, more matter arrived to the inner disc, and accretion switched to the unstable regime 
  where matter penetrates through the magnetosphere in the unstable ``tongues'' (e.g., \citealt{RomanovaEtAl2008,KulkarniRomanova2008,KulkarniRomanova2009}). The onset of the unstable regime 
  depends on the effective gravity (the sum of the gravitational and centrifugal potential), and therefore 
  depends on the   ratio $\bar{r}_m/r_c$.  
  According to \citet{BlinovaEtAl2016}, 
accretion becomes unstable, if $r_m/r_c\lesssim 0.71$ (in their set of simulations, where the magnetic axis is tilted by  
 $\theta=5^\circ$).

  In our model the star rotates slowly compared with the inner disc, $\bar{r}_m/r_c\approx 0.24$. However,
  accretion is stable up to $t\approx 60$, and becomes unstable at $t>60$. At $t<60$, the magnetospheric radius was only slightly larger during stable regime.
 We conclude that in the case of the tilted rotational axis the unstable regime is less favorable compared with the aligned case considered  by \citet{BlinovaEtAl2016}.

\subsection{Both the rotational and magnetic axes are tilted: $\beta=20^\circ$, $\theta=20^\circ$ (Model B)}
\label{sec:B}

Next, we consider the model where both axes are misaligned. 
In this model, the mass of the disc is $\sim 3$ times smaller than that in Model A. The bottom panels of Fig. \ref{2d-initial-8} show that we start from a thin disc, which expands and becomes comparable in thickness with the disc in Model A. The density in the disc is $\sim 3$ times smaller than in Model B. 

The  overall evolution of the disc is similar to
that in Model A. 
Namely, initially, the
inner parts of the disc are warped,  then tilted, and precess about the rotational axis of the star.
 After 1-2 periods of precession the disc settles near the rotational equatorial plane of the star, and the
disc normal has a small tilt angle, $\beta_t\sim 5^\circ-10^\circ$ relative to the rotational axis of the star (see Fig. \ref{3d-t20-bet20-16}).  

 In this model, the disc is of the lower density and as a result 
the time-averaged radius of the magnetosphere 
 $\bar{r}_m\approx 4$ is larger than in Model A ($\bar{r}_m\approx 3.4$).
 The radius of the tilted disc is slightly smaller than that in Model A:
 $r_t\approx 21.8$.  The disc is mainly flat, but, compared with Model A, there is an additional wavy structure
 connected with rotation of the magnetic axis about the rotational axis of the star.
 The alignment of the inner disc normal with the rotational axis of the star occurs faster than in Model A. This may  be due to the lower density  in the disc.  
Namely, in Eq. \ref{eq:Gamma_w} the warping rate $\Gamma_w$ is inversely proportional to the surface density  $\Sigma$,
and this may explain the faster variation of the tilt angle in Model B. 

We calculated the accretion rate onto the star. 
We note that the magnetic moment of the star is tilted
about the disc normal at different angles. During
one rotational period, the position of the magnetic axis
relative to the initial disc axis varies between strongly tilted
  ($\chi=\beta+\theta=40^\circ$)   and the aligned one ($\chi=\theta-\beta=0^\circ$). 
In the former case,  the accretion through funnel streams is more favorable due to the high tilt of the magnetosphere towards the disc. 
This  leads to the variation of the accretion rate at the surface of the star.  The right panel of Fig. \ref{pm-all-2} 
shows several maxima and minima which  correspond
 to different tilts of the magnetic axis relative to the disc.

\begin{figure*}
\centering
\includegraphics[width=16cm]{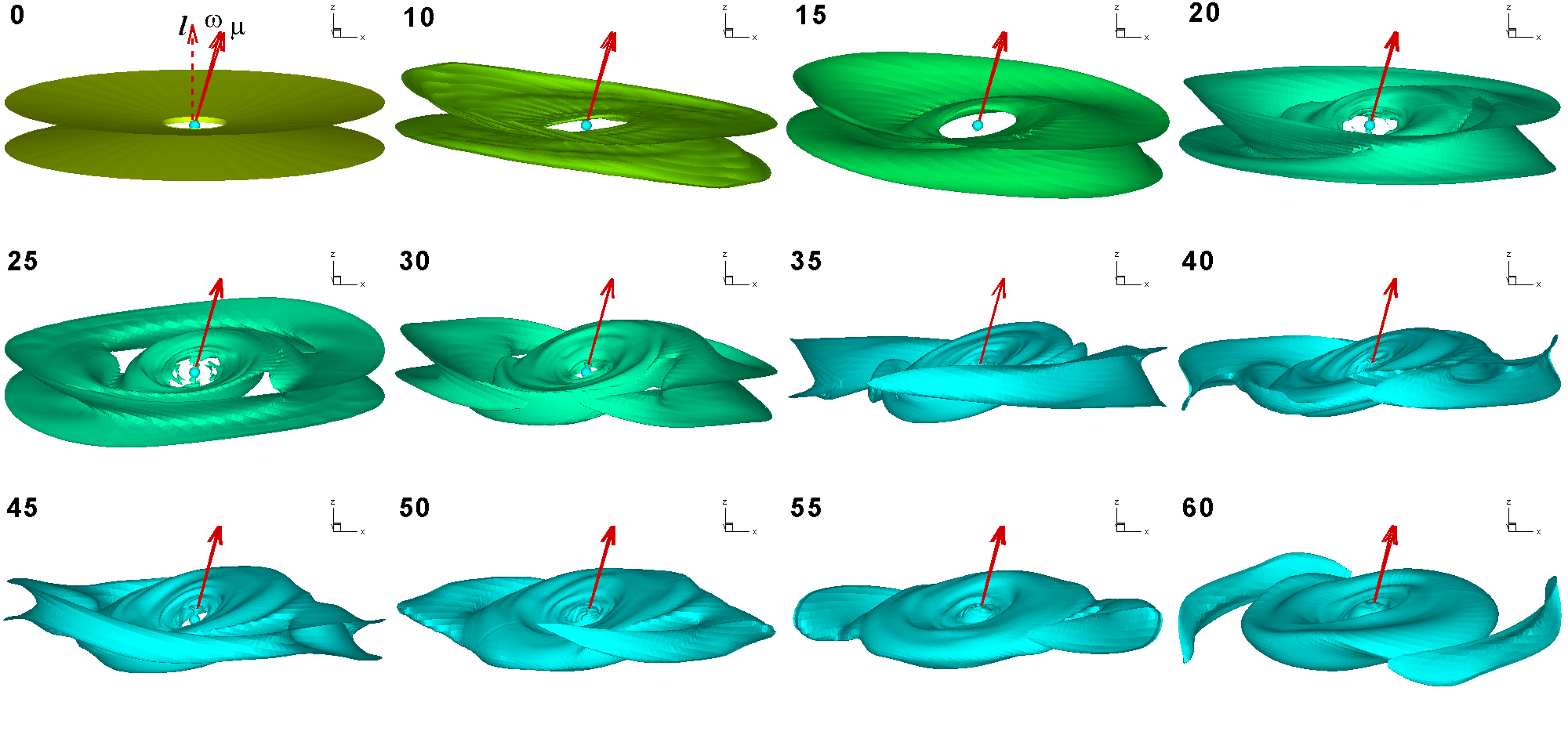}
\caption{3D views of the disc in Model D at different moments of time.
The color background shows density levels which vary from 
$\rho=0.3-1.5$ in the top row of panels, to $\rho=0.15-0.2$ in two bottom rows of panels. 
} \label{3d-d05-c2-3-12}
\end{figure*}

 We chose two moments in time corresponding to the maximum ($t=22$) and minimum ($t=28$) of the accretion rate and checked the position of the magnetosphere and the nature of the
  matter flow at these moments.   
  The  top left  panel of Fig. \ref{2d-3d-min-max} shows that at $t=22$, the magnetic axis $\bmu$ is strongly tilted about the rotational axis of the inner disc, and two funnel streams are formed. The bottom left panel shows that at $t=28$ the magnetic axis is almost perpendicular to the disc, accretion in funnels is less favorable, and only weak funnel streams formed. 
Middle and right panels of Fig. \ref{2d-3d-min-max}
show the $x'z'$  and $y'z'-$slices of density distribution
 during high and low tilts of
the magnetic axis. One can see that funnels form more efficiently
during episodes of higher tilt of the magnetosphere. 

The amplitudes of maxima in the curve for the accretion rate are larger initially when the disc normal had a higher tilt about the rotational axis of the star.
Later, when the disc becomes almost aligned, the tilt of the magnetic axis only slightly varied about the normal to the disc and the
amplitudes of maxima become smaller. 

 This model shows that in the case when both axes are misaligned,  the main result is similar to that in the case of the aligned dipole: 
 the disc tends to be in the rotational equatorial plane of the star. 
We consider possible explanations 
of the disc alignment in Sec. \ref{sec:alignment}.

\subsection{Dependence on the disc mass and rotation rate (models A1, B1, C)}
\label{sec:diff}

The supplement models A1 and B1 are identical to models A and B, but the disc mass is $\sim 3$ times larger/smaller, respectively. Simulations have shown the same main result: the tilted disc 
settled approximately in the equatorial plane of the star such that the normal to the inner disc is tilted only at a small angle relative to the rotational axis of the star. The accretion rate is 3 times smaller/larger, respectively.  In Model B1,  the variability in the matter flux, associated with different tilts of the magnetic axis has also been observed. Episodes of unstable accretion were observed in Model B1, where the magnetospheric radius is smaller. These models have shown that result does not depend on the factor of 3 variations in the disc mass.

 We also tested a model similar to Model A, but for a faster rotating star, $r_c=8.6$, $P_\star=5.2$ (Model C).  We observed the formation of the inner tilted disc similar to that in Model A. 
 However, the normal to the inner disc is tilted at a slightly larger angle:  $\beta_n\sim 15^\circ-20^\circ$. We suggest that there may be a dependence of the tilt on the rotation rate of the star. 
 We note that the physics of the disc-magnetosphere interaction often depends on the ratio $r_m/r_c$ (e.g., \citealt{GhoshLamb1978,BlinovaEtAl2016}). 
 This ratio $\bar{r}_m/r_c\approx 0.39$ is larger in this model versus models A and B: $\bar{r}_m/r_c\approx 0.24, 0.28$ (see Tab.  \ref{tab:models}). We further investigate this issue in Sec. \ref{sec:DEF}.

\section{Models of tilted discs (D, E, F)}
\label{sec:DEF}

In this section, we consider discs that show a large tilt angle at the end of simulations. 
In these models, we placed the initial radius of the disc closer to the star, at $r_{\rm in}=5.7$ (versus 8.6 in the above models) and therefore a stronger dipole magnetic field threads the disc.  We also took  faster rotating stars. We observed  qualitatively different result: 
the normal to the  inner disc was systematically tilted at a large angle away from the rotational axis of the star. 

In these models, the tilt of the rotational axis is $\beta=15^\circ$, and tilts of the magnetic axes are $\theta=2^\circ$ or $15^\circ$. The corotation radius $r_c=8.6$ or $r_c=5.1$ which correspond to periods of the star $P_\star=5.2$ and 
$P_\star=2.4$
We took smaller values of the magnetic moment: $\mu'=0.5$ and $0.3$. See Tab. 1 for all set of parameters.
We show sample results for these models.

Three left panels of Fig. \ref{2d-d05-d03-8} show $x'z'-$slices of the initial density distribution and sample magnetic field lines in models D and F at times $t=0, 15, 20$. Right panels show close view of the magnetospheric accretion at $t=20$. Note that the magnetospheric radii are smaller than in models $\rm A-C$:
$\bar{r}_m=2.9$ and $2.1$ in models D and F, respectively. 
 
 Fig. \ref{3d-d05-c2-3-12} shows 3D views of the disc in Model D. One can see  that the inner disc becomes warped, then tilted, and the inner disc seems to be disconnected from the outer parts of the disc. The radius of the tilted disc 
 $r_t\approx 19$ and its normal vector is tilted away from the rotational axis of the star at an angle $\beta_t\approx 30^\circ-40^\circ$, which is much larger than that in models $\rm A-C$. We observed very slow precession in this model.
  
 In two other models  (E and F) similar tilted discs were formed, with tilt angles, $\beta_t\approx 30^\circ-40^\circ$,  and tilt radii 
  $r_t=19.4$ and $17.7$.  Discs in models E and F precess with usual rates of
  1-1.5 precession periods per simulation run.   
 In Model D the precession is very slow.

The magnetic field lines wrap due to the rotation of the inner disc and rotation of the star.   The right-hand panel of Fig. \ref{3d-mag-inflation} shows the field lines in Model D during a relatively early time of evolution ($t=30$). One can see that the field lines wrap about the disc normal because the disc rotates more rapidly than the star. However,  wrapping about the rotational axis of the star was also observed. On the longer time scale, the field lines form a magnetic tower about the rotational axis of the star.

\begin{figure*}
\centering
\includegraphics[width=16.0cm]{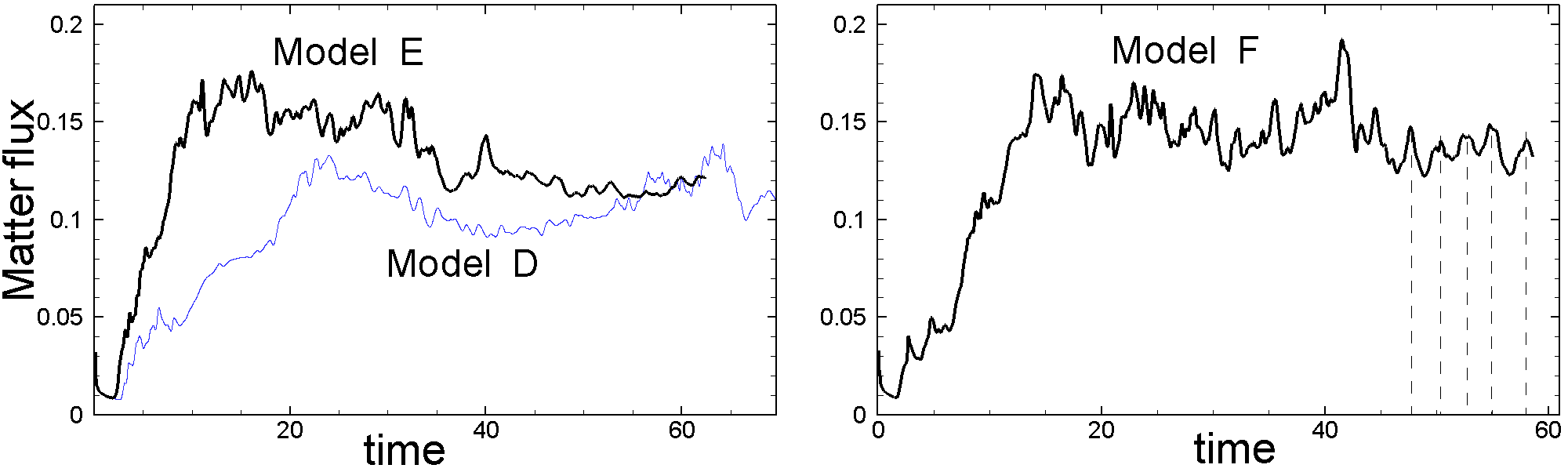}
\caption{Temporal variation of matter flux $\dot M$ at the surface of the star.  Left-hand panel shows $\dot M$ in models  D (blue color) and E, and right-hand panel for Model  F. Vertical dashed lines show maxima corresponding to high tilt of the magnetic axis towards the inner disc.}\label{pm-d05-d03}
\end{figure*}

Fig. \ref{pm-d05-d03} shows matter fluxes in these models.  In model  F, where $\theta=15^\circ$, one of the variabilities is connected with different tilts of the magnetic axis relative to the disc (like in Model B, see Fig. \ref{pm-all-2}). The quasi-period of variability approximately equals to the period of the star, $P_\star=2.4$). Variabilities in models E and D and the flaring component of variability in Model F are connected with non-stationary and/or unstable accretion.\footnote{Note that in stars with the smaller magnetosphere, the unstable regime occurs more easily than in stars with larger magnetospheres \citep{BlinovaEtAl2016}.}
  
One of the main differences between this set of models and  the earlier discussed set of models ($\rm A-C$) is that 
the inner disc was initially closer to the star, and 
 stronger dipole field threads the disc. Therefore, the role of the dipole component 
(which helps to tilt the disc) is more significant. 
 Namely, in Eq. \ref{eq:Bz}, the dipole components $B_\phi$ and $B_z$ are important in providing the force   $F_z(r)$  and warping torque $N_w(r)$ which persistently tilt the disc away from the equatorial plane of the star.
\footnote{In the real situation, the tilt may depend on the diffusivity at the disc-magnetosphere boundary and the level of penetration of the stellar field to the inner parts of the disc.} On the other hand,  we noticed in the test Model C and current models, that the disc is more tilted when the star rotates more rapidly. We calculated the  ratios $\bar{r}_m/r_c$ and noticed that they are larger than in models $\rm{A-C}$ (see Tab. 1).
 We discuss  possible reasons which lead to alignment or tilting of the disc in the next section.
 
\section{Mechanisms of disc alignment and tilting}
\label{sec:alignment}

In our models the disk breaks up into two parts. The tilt of the inner disc, $\beta_t$, is different in different models 
Below we discuss possible mechanisms explaining different tilts of the inner disc. 

\subsection{Mechanisms of disc alignment}

In models $\rm{A-C}$ the normal to the tilted disc tends to align with the rotational axis of the star. 
Below we consider two possible explanations for this phenomenon.

\begin{figure*}
\centering
\includegraphics[width=8cm]{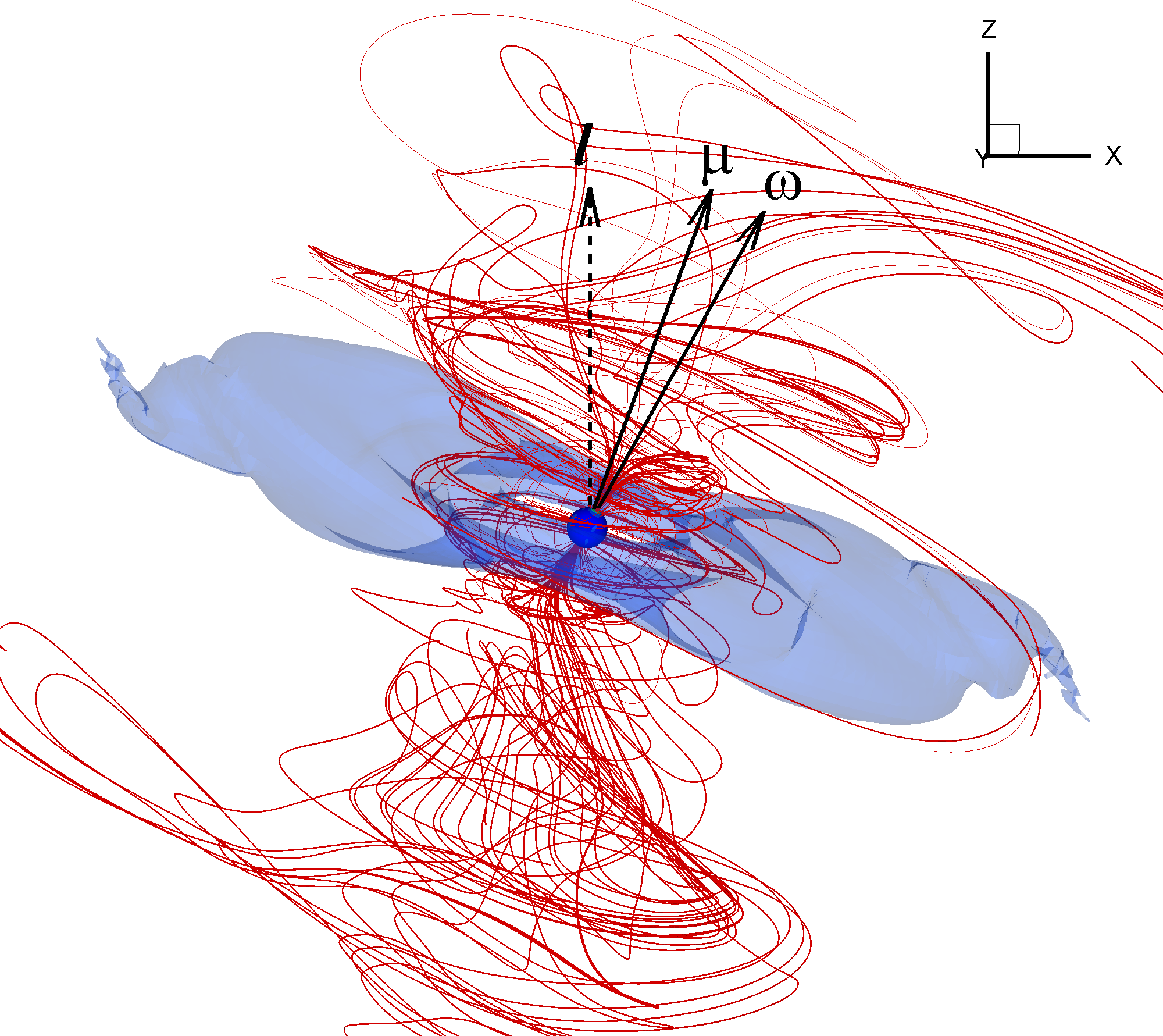}
\includegraphics[width=8cm]{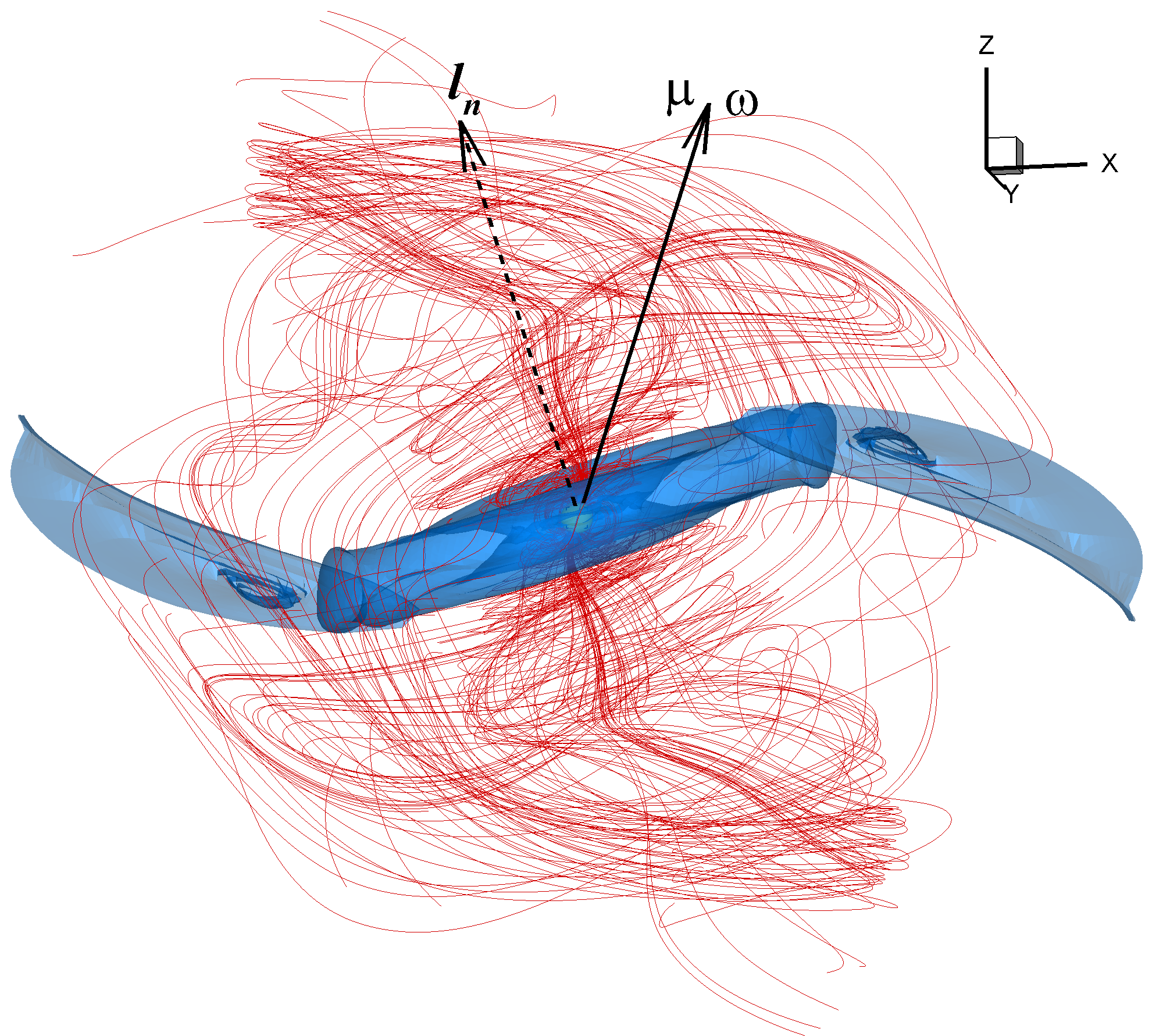}
\caption{{\it Left panel:} Figure demonstrates inflation of the magnetic field lines
in Model B at $t=62$.  The color background
shows one of the density levels,  $\rho=0.17$. Lines are sample magnetic field
lines. {\it Right panel:} Same but for Model D at 
$t=30$ and for density level $\rho=0.3$. 
}\label{3d-mag-inflation}
\end{figure*}

In our models, the magnetic field lines are wrapped due to the differential rotation of their foot-points connecting the star and the disc. 
The inner disc rotates more rapidly than the magnetosphere of the star, and the field lines are wrapped
 about the normal to the inner disc and expand forming a local magnetic tower. The inner disc changes its tilt and the tower changes its direction.  At the same time, a star rotates and the field lines wrap about the rotational axis of the star. On a long time scale and larger spatial scales, the magnetic tower becomes more and more symmetric about the rotational axis of the star.  Left-hand panel of Fig. \ref{3d-mag-inflation} shows the magnetic tower
observed in  Model B at $t=62$. Note that at this time the normal to the inner disc has a small angle relative to the rotational axis of the star, which makes the tower more symmetric.    

The right-hand panel of Fig. \ref{3d-mag-inflation} shows the tower in Model D, where the disc normal is tilted at a large angle
and at the earlier moment in time, $t=30$. One can see that near the disc the field lines wrap about the disc normal, while at larger distances the wrapping about the stellar rotational axis is seen. In reality, both components of the wrapped field are present in all models. The azimuthal component of the field above and below the disc can be presented as a sum of the field wrapped about the disc normal (marked with letter $d$) and  stellar rotational axis (marked with letter $s$): 
$B_{\phi}^{top}=B_{\phi}^d+B_{\phi}^{s,top}$ and $B_{\phi}^{bot}=B_{\phi}^d+B_{\phi}^{s,bot}$. The disc components of the field are approximately equal. However, the stellar component is expected to be  stronger near parts of the disc that are closer to the rotational axis of the star. In the right panel of Fig. \ref{3d-mag-inflation}, the field is stronger near the top right and bottom left parts of the tilted disc. Therefore, there is the magnetic force acting on the disc which is the difference between magnetic pressure at the top and bottom sides of the disc:
\begin{equation}
F_{z}(r,\phi)={1\over 8\pi}\left[
(B_{\phi}^d+B_{\phi}^{s,top})^2- (B_{\phi}^d+B_{\phi}^{s,bot})^2\right] .
\label{eq:Bz-wrap}
\end{equation} 
The corresponding  torque acts to align the normal to the disc with the rotational axis of the star.
This torque acts in the direction opposite to the warping torque. 
We suggest that this may be a possible mechanism for the disc alignment in models $\rm A-C$.
 Note that in Eq. \ref{eq:Bz-wrap}, the magnetic pressure results from the winding of the field lines, threading the disc. Note that in Eq. \ref{eq:Bz} for magnetic force providing the warping torque, the azimuthal field associated with the inflated field is taken to be equal on the top and bottom sides of the disc, and the main asymmetry is connected with the $B_\phi-$component of the dipole field. We suggest that in models $\rm A-C$ (where the inner disc was located at a larger distance from the star), the dipole component has been relatively weak, and the alignment torque dominates over warping torque. In opposite, in models D, E, F (where the inner disc was closer to the star), the dipole component is stronger, and warping torque dominates. 
 
\begin{figure*}
\centering
\includegraphics[width=17cm]{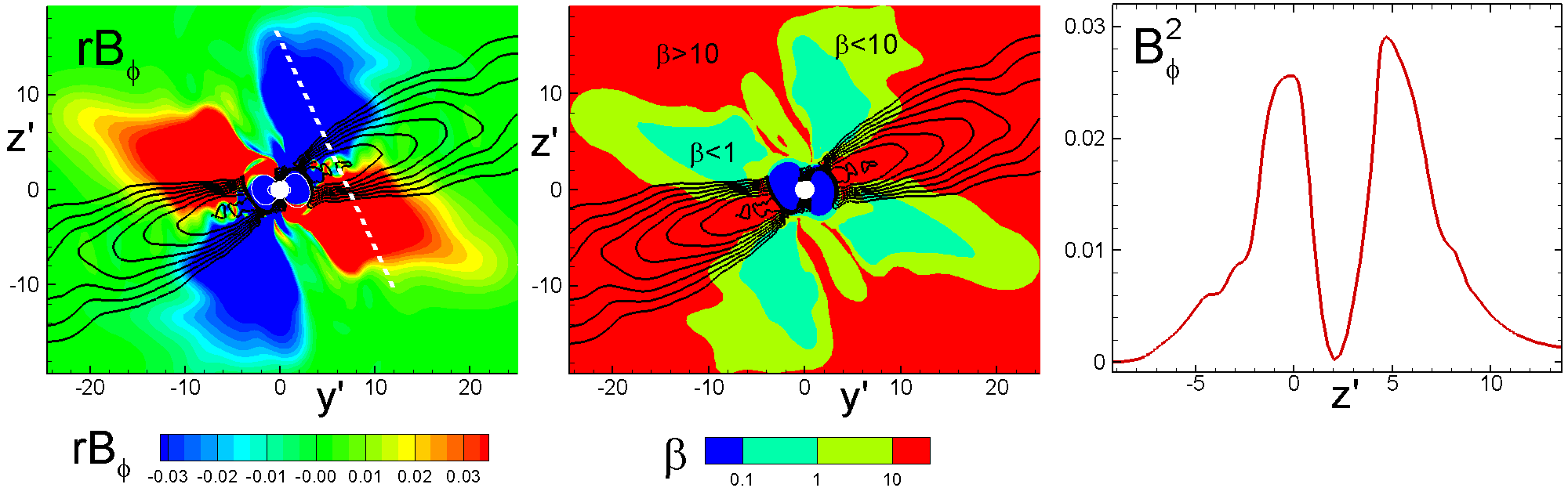}
\caption{\textit{Left panel:} the 
distribution of the poloidal current $rB_\phi$  in the slice $y'z'$ in Model A at $t=50$. \textit{Middle panel:} the distribution
 of the plasma parameter  $\beta=8\pi p/B^2$. We show $\beta$ in three regions, separated by values  $\beta=0.1, 1, 10$.   
 \textit{Right panel:} the distribution of the $B_\phi^2$ along the dashed line shown in the left panel. }\label{current-3}
\end{figure*}

To investigate further this issue, we calculated the poloidal current $J_p\sim r B_\phi$ and observed that the current flows above and below the disc, and it is almost symmetric about the disc plane (see left panel of Fig. \ref{current-3}). 
We draw a line perpendicular to the disc (see white dashed line in the left panel) and calculated the value of ${B_\phi}^2$ along this line. The right panel of Fig. \ref{current-3}  shows that the magnetic pressure distribution is not perfectly symmetric about the plane of the disc, and the pressure difference provides the magnetic force, which may be responsible for the tilting of the disc. Note that the magnetic pressure dominates over the matter pressure in the corona above and below the inner parts of the disc.  Middle panel of Fig. \ref{current-3} shows the distribution of the plasma parameter $\beta=8\pi p/B^2$. One can see that $\beta<1$ in the corona above and below the disc at $r\lesssim 11$ (see darker green regions). There is also a region where the matter pressure dominates, but the magnetic pressure is still significant and can contribute to the dynamics of the disc (see the light-green region at $r\lesssim 14$ where $\beta<10$). 
The sizes of these regions vary in time and also from model to model. However, they are always a few times larger than the magnetospheric radius $r_m$.  
We suggest that this magnetic force and corresponding torque may drive the tilted discs towards the aligned position.

Another possible explanation for the disc alignment is the magnetic Bardeen-Petterson effect (see Sec. \ref{sec:BP}), where the viscous and precession torques
push the inner disc to be aligned with the equatorial plane of the star.   
 Typically, we observe 1-2 periods of precession.
This time may be too short for the development of the Bardeen-Petterson effect. In the case of compact stars, the time scale to
achieve the Bardeen-Petterson alignment is different in different models and varies from 
a few precession time scales (evaluated at $R_{\rm BP}$) up to $10-100$  (e.g., \citealt{Pringle1992}).

\subsection{Mechanisms of tilting}
    
The warping instability discussed in Sec. \ref{sec:warping} always acts to tilt the inner disc normal away
from the rotational axis of the star (see, e.g., \citealt{LaiEtAl2011}). 
The warping torque operates at distances comparable with the magnetospheric radius. It is stronger 
 in models D, E, F where the inner disc is closer to the star, and a stronger dipole field threads the inner disc.

On the other hand, the winding of the field lines about the stellar rotational axis provides a force that acts to align the disc.
If a star rotates slowly (as in models A, A1, B, B1) then the role of winding is more significant, because 
there is a larger difference between angular velocities of the star and the disc. In these models, we observe almost aligned discs. If a star rotates more rapidly (like in models $\rm C-F$) then the role  of the force associated with winding
is less important, and the warping force dominates.

\section{Discussion and conclusions}
\label{sec:discussion}

We performed three-dimensional MHD simulations of
accretion onto a rotating magnetized star where both
the magnetic and rotational axes of the star are tilted about the
rotational axis of the disc. 

\subsection{Summary: Dependence on parameters}

Our simulations are exploratory and are aimed at understanding the matter flow 
near the magnetized star where both the magnetic and rotational axes are tilted. We varied different parameters
 (see Tab. \ref{tab:models}). 
In addition to evident parameters, such as the magnetic moment, $\mu'$, or period of the star, $P_\star$, 
we also varied the initial position of the inner disc, $r_{\rm in}$, and observed strong dependence on this parameter.   Below, we conclude about dependence on different parameters.

\begin{itemize}

\item $r_{\rm in}-$the initial positions of the inner disc. We observed that in models $\rm A-C$, where the inner disc is located at a larger radius, $r_{\rm in}=8.6 R_\star$, the final tilt angle of the disc is smaller, $\beta_t\sim 5^\circ-10^\circ$, compared with models D, E, F, where the disc is located closer to the star,  $r_{\rm in}=5.7 R_\star$, and the tilt of the disc is larger, $\beta_t\sim 30^\circ-40^\circ$. 

\item $\beta-$initial tilt of the rotational axis of the star relative  to the disc normal. We did not see a difference between results in models with $\beta=20^\circ$ and $\beta=15^\circ$.

\item $\theta-$the tilt of the magnetic axis relative to the rotational axis. There is almost no difference in results for models with almost aligned ($\theta=2^\circ$) and misaligned ($\theta=20^\circ$, $\theta=15^\circ$) cases. The main difference is that in models with larger $\theta$, we observed variability in the matter flux, which is  
 associated with different tilt angles  between the magnetosphere and the disc. 

\item $r_c-$the corotation radius and $P_\star-$period of the star. In models D, E, F, stars rotate more rapidly than in models A, A1, B, and B1, and this could be a factor that leads to larger tilts of discs in these models. We suggest that at smaller values of $r_c$ and larger values of $\bar{r}_m/r_c$, the difference in angular velocities between the star and the disc is smaller, and winding of the field lines (which helps to align the disc) is less efficient.

\item $\mu'-$the magnetic moment of the star: $\mu'=1, 0.5, 0.3$. We observed that the size of the tilted disc, $r_t$, decreases with $\mu'$. This is an expected result, because at smaller values of $\mu'$ the magnetic force is smaller.
 
\item $M_d-$mass of the disc. In test simulations with $\sim 3$ times lower disc mass (models B and A1) we observed similar parameters for tilted discs. However, discs were warped and tilted more rapidly. 
 This may be explained by the fact that the warping rate has an inverse dependence on the surface density: $\Gamma_w\sim \Sigma^{-1}$ (see Eq. \ref{eq:Gamma_w}). 
 
\item $\tau_{\rm sim}-$duration of simulations.  Originally, we included into consideration only the longest simulation runs (models A and B) which show almost aligned discs. However, later, we realized that models D, E, F are also valuable because they show persistent tilts. In these models, the time measured in Keplerian rotations at, $\tau_{\rm sim}/P_0$,
is shorter. However, time measured in periods of stellar rotation, $\tau_{\rm sim}/P_\star$, is comparable or longer  
than in models A and B. 
The rotation of the star is an important factor in winding the field lines and may influence the physics of the process.

\end{itemize}

\subsection{Conclusions}
 
\noindent\textbf{1.}  Simulations show that the disc-magnetosphere interaction led to the formation of tilted, almost flat discs in all models. However, discs may have different tilts. The tilt angles of the disc normal relative to the rotational axis of the star are small ($\beta_t\sim 5^\circ-10^\circ$) in models, where the star rotates slowly and where initially the disc is located at a larger distance from the star so that a weaker dipole field threads the disc. When stars rotate more rapidly and the inner disc is located closer to the star (so that the stronger dipole field threads the disc), the tilt angles are larger ($\beta_t\sim 30^\circ-40^\circ$).

\smallskip

\noindent\textbf{2.} The sizes of the tilted discs systematically increase with the strength of the magnetic field, $\mu'$.
They vary in the range of 
 $r_t \approx 17.7 - 24.3$ if measured in stellar radii. They are typically  
 $\sim 5.3-8.6$ times larger than the magnetospheric radii.

\smallskip

\noindent\textbf{3.} Tilted discs slowly precess in most of models.  The time scale of precession is $\tau_p\sim 50 P_0$, where $P_0$ is the period of Keplerian rotation at $r=R_0\approx 2.86 R_\star$. 

\smallskip

\noindent\textbf{4.} In models with a significant  tilt of the magnetic axis ($\theta=20^\circ$ and $15^\circ$), the accretion rate onto the star varied due to the different positions of the magnetospheric axis about the inner disc.
Accretion is more favorable when the magnetic axis is strongly tilted towards the disc plane. 
The quasi-period of variations is close to the period of the star.   

\smallskip

\noindent
{5.} Accretion in the unstable regime has been observed in models with higher-mass discs and  smaller tilts of the magnetosphere.

\smallskip

Overall, tilted discs are expected to form around magnetized stars with the tilted rotational axis. However, the tilt angle and other parameters of the disc depend on the properties of the star and details of the disc-magnetosphere interaction.

\subsection{Application to different stars}

Tilted precessing discs are expected in different types of accreting magnetized stars.

\smallskip

1. The signs of tilted discs are observed in cataclysmic variables (CVs). They are often observed as temporary features. The origin of the tilt is not well understood  (see, e.g., \citealt{MontgomeryMartin2010}).  
\footnote{\citet{FateevaEtAl2016} studied accretion onto magnetized stars with the tilted rotational axis  in 3D MHD simulations in application to intermediate polars. However, only a small (a few per cent) temporary tilts of the inner disc were observed in these simulations.}
We suggest that tilted discs may result from the action of the magnetic field, as observed in our models,
where the disc is expected to be tilted as long as the dipole magnetic field partly threads the disc. 
Note that the inflated field lines may drive outflows or jets from the disc-magnetosphere boundary. The orientation of the magnetic tower
 may be important for determining the direction of such outflows (e.g., \citealt{LovelaceEtAl2014}). 

\smallskip

2. In another example, accreting millisecond pulsars in X-ray binaries show a variety of quasi-periodic oscillations (QPOs)  (see, e.g., \citealt{Vanderklis2006}).  
We suggest that the low-frequency QPOs can be connected with the precession of tilted discs, driven by the magnetic forces (see also \citealt{Lai1999}).

\smallskip

3. The long-term variabilities of unknown nature have been observed in classical T Tauri stars and Ae Herbig stars (e.g., \citealt{ArtemenkoEtAl2010,RigonEtAl2017}). Some of them may be connected with tilted precessing discs.

\section{Data availability}

3D and 2D plots shown in the paper were produced using data obtained in 3D MHD simulations. These data will be 
shared on reasonable request to the corresponding author.

\section*{Acknowledgments}

Authors thank anonymous referee for insightful comments.  Resources supporting this
work were provided by the NASA High-End Computing (HEC) Program
through the NASA Advanced Supercomputing (NAS) Division at Ames
Research Center and the NASA Center for Computational Sciences
(NCCS) at Goddard Space Flight Center. 
MMR and RVEL were supported in part by the NSF grant AST-2009820.

\end{document}